%% file: main.tex
\documentclass[journal=jctcce,manuscript=article]{achemso}

\usepackage{achemso}
\setkeys{acs}{
    abbreviations = true,
    maxauthors = 10
}
\usepackage{graphicx} 
\usepackage{chemformula} 
\usepackage[T1]{fontenc} 
\usepackage{amssymb}
\usepackage{booktabs}
\usepackage[acronym]{glossaries}

\usepackage{caption}
\usepackage{subcaption}
\usepackage{rotating}

\usepackage{tabu}
\usepackage[capitalise,noabbrev]{cleveref}
\usepackage{siunitx}

\DeclareSIUnit{\calorie}{cal}
\DeclareSIUnit{\Calorie}{\kilo\calorie}
\DeclareSIUnit\atom{atom}
\DeclareSIUnit\step{step}
\DeclareSIUnit\angstrom{\text {Å}}

\usepackage[version=4]{mhchem} 

\usepackage{xcolor}
\definecolor{mygreen}{RGB}{28,172,0}
\definecolor{mylilas}{RGB}{170,55,241}
\definecolor{codegreen}{rgb}{0,0.6,0}
\definecolor{codegray}{RGB}{105,105,105}
\definecolor{codepurple}{rgb}{0.58,0,0.82}
\definecolor{backcolour}{RGB}{240,240,240}

\usepackage{listings}
\newcommand\YAMLcolonstyle{\color{red}\mdseries}
\newcommand\YAMLkeystyle{\color{black}\ttfamily\footnotesize}
\newcommand\YAMLvaluestyle{\color{blue}\mdseries}

\makeatletter

\newcommand\language@yaml{yaml}

\expandafter\expandafter\expandafter\lstdefinelanguage
\expandafter{\language@yaml}
{
  keywords={true,false,null,y,n},
  keywordstyle=\color{darkgray}\bfseries,
  basicstyle=\YAMLkeystyle,                                 
  sensitive=false,
  comment=[l]{\#},
  morecomment=[s]{/*}{*/},
  commentstyle=\color{purple}\ttfamily,
  stringstyle=\YAMLvaluestyle\ttfamily,
  moredelim=[l][\color{orange}]{\&},
  moredelim=[l][\color{magenta}]{*},
  moredelim=**[il][\YAMLcolonstyle{:}\YAMLvaluestyle]{:},   
  morestring=[b]',
  morestring=[b]",
  literate =    {---}{{\ProcessThreeDashes}}3
                {>}{{\textcolor{red}\textgreater}}1     
                {|}{{\textcolor{red}\textbar}}1 
                {\ -\ }{{\mdseries\ -\ }}3,
}

\lst@AddToHook{EveryLine}{\ifx\lst@language\language@yaml\YAMLkeystyle\fi}
\makeatother

\newcommand\ProcessThreeDashes{\llap{\color{cyan}\mdseries-{-}-}}

\DeclareMathOperator*{\argmaxA}{arg\,max}

\lstdefinestyle{mystyle}{ 
    backgroundcolor=\color{backcolour},   
    commentstyle=\color{codegreen},
    keywordstyle=\color{magenta},
    numberstyle=\tiny\color{codegray},
    stringstyle=\color{codepurple},
    basicstyle=\ttfamily\footnotesize,
    breakatwhitespace=false,         
    breaklines=true,                 
    captionpos=b,                    
    keepspaces=true,                 
    numbers=left,                    
    numbersep=5pt,                  
    showspaces=false,                
    showstringspaces=false,
    showtabs=false,                  
    tabsize=2
}

\SectionNumbersOn

\author{Moritz R. Sch\"{a}fer}
\affiliation[TheoChem]{Institute for Theoretical Chemistry, University of Stuttgart, Pfaffenwaldring 55, 70569 Stuttgart, Germany}

\author{Johannes K\"{a}stner}
\affiliation[TheoChem]{Institute for Theoretical Chemistry, University of Stuttgart, Pfaffenwaldring 55, 70569 Stuttgart, Germany}
\email{kaestner@theochem.uni-stuttgart.de}

\title[Enhanced Representation Based Sampling]
  {Enhanced Representation-Based Sampling for the Efficient Generation of Datasets for Machine-Learned Interatomic Potentials}

\begin{document}

\begin{tocentry}
    \includegraphics[width=\linewidth]{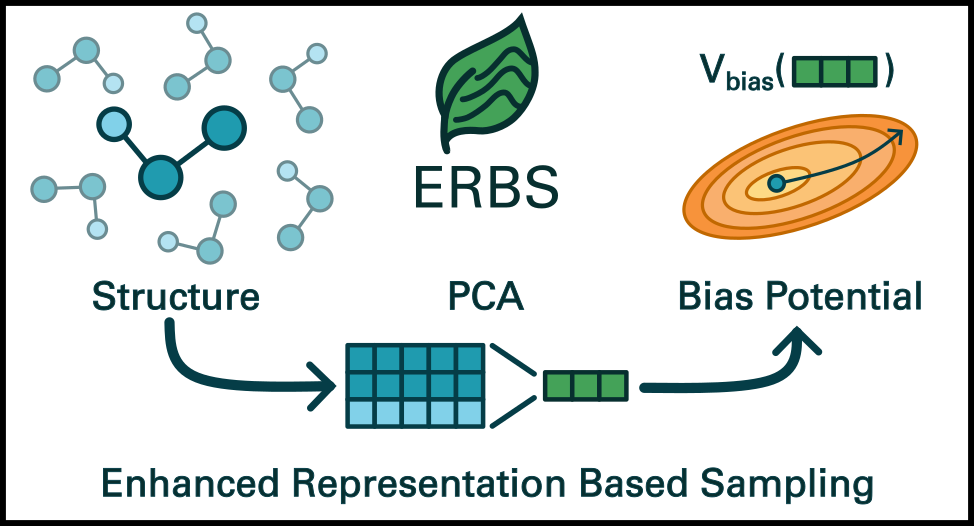}
\end{tocentry}

\include{abstract}
\include{body}

\include{supplementary-information}

\bibliography{references}

\end{document}

%% file: abstract.tex
\begin{abstract}

In this work, we present Enhanced Representation-Based Sampling (ERBS), a novel enhanced sampling method designed to generate structurally diverse training datasets for machine-learned interatomic potentials.
ERBS automatically identifies collective variables by dimensionality reduction of atomic descriptors and applies a bias potential inspired by the On-the-Fly Probability Enhanced Sampling framework.
We highlight the ability of Gaussian moment descriptors to capture collective molecular motions and explore the impact of biasing parameters using alanine dipeptide as a benchmark system.
We show that free energy surfaces can be reconstructed with high fidelity using only short biased trajectories as training data.
Further, we apply the method to the iterative construction of a liquid water dataset and compare the quality of simulated self-diffusion coefficients for models trained with molecular dynamics and ERBS data.
Further, we active-learn models for liquid water with and without enhanced sampling and compare the quality of simulated self-diffusion coefficients.
The self-diffusion coefficients closely match those simulated with a reference model at a significantly reduced dataset size.
Finally, we compare the sampling behaviour of enhanced sampling methods by benchmarking the mean squared displacements of \ce{BMIM+BF4-} trajectories simulated with uncertainty-driven dynamics and ERBS and find that the latter significantly increases the exploration of configurational space.

\end{abstract}

%% file: body.tex
\section{Introduction}

Machine-learned interatomic potentials (MLIPs) have proven themselves to be a suitable approach to studying the dynamics of atomistic systems over the last couple of years\cite{unkeMachineLearningForce2021,behlerMachineLearningPotentials2021}.
By training a machine learning model on the results of ab initio calculations, it is possible to perform molecular dynamics simulations with nearly the accuracy of the reference methods and cost that scales linearly with system size.
Following the seminal work by Behler and Parinello\cite{behlerGeneralizedNeuralNetworkRepresentation2007a,Behler2010} and Bartok and Csanyi\cite{bartok09a,bartok10a,bartok13a}, numerous model architectures have been proposed.
These include a broad range of approaches, from descriptor-based kernel models\cite{chmielaMachineLearningAccurate2017,drautzAtomicClusterExpansion2019} to equivariant message passing models\cite{batatiaMACEHigherOrder2022,batznerE3equivariantGraphNeural2022} and many others\cite{schuttSchNetDeepLearning2018a, wangDeePMDkitDeepLearning2018,dullingerAcceleratingMolecularDynamics2024}.
While a significant focus by the community has been put into developing new architectures, all data-driven models are only as good as the data they were trained on.
The quality of training data is becoming increasingly critical with the emerging interest in atomistic foundation models\cite{batatiaFoundationModelAtomistic2025,eastmanSPICEDatasetDruglike2023a}.
Unlike potentials trained for a single system, these models aim to maximize transferability and robustness across domains\cite{zillsMLIPXMachinelearnedInteratomic2025}.
However, current observations suggest inconsistent performance away from equilibrium and in the treatment of soft modes\cite{dengSystematicSofteningUniversal2025}, likely caused by a heavy reliance on relaxation trajectories as training data.
Consequently, the development of methods capable of generating structurally diverse datasets is of utmost importance for the continued advancement of general-purpose atomistic models.

Historically, constructing datasets for MLIPs was a costly and labour-intensive process, as reference data was generated using ab initio molecular dynamics\cite{chmielaMachineLearningAccurate2017,unkeMachineLearningForce2021}.
Sampling a configuration space with this approach results in highly correlated samples, each requiring costly calculations.
With the emergence of active learning approaches\cite{smithANI1ExtensibleNeural2017}, researchers have started to use preliminary MLIPs for the sampling of candidate configurations, for example, by molecular dynamics.
Thus, it became possible to reduce the usage of costly quantum chemical reference methods only on the most informative candidates, resulting in more compact and more informative datasets.

A key advancement was the introduction of uncertainty estimation methods to the realm of MLIPs.
With techniques such as model ensembling\cite{smithANI1ExtensibleNeural2017,buskGraphNeuralNetwork2023a,thalerScalableBayesianUncertainty2023}, Gaussian processes\cite{guanConstructionReactivePotential2018,vandermauseFlyActiveLearning2020}, or optimal experimental design\cite{zaverkinExplorationTransferableUniformly2021}, it became possible to estimate the error of model predictions.
Using these uncertainty estimates, MLIP-driven simulations can be terminated when uncertainty is high, and the most uncertain configurations can be selected for recalculation using ab initio methods, an approach called uncertainty-driven dynamics (UDD).
The model is then retrained on the updated dataset.
This iterative process, often called active learning or learning-on-the-fly\cite{vandermauseFlyActiveLearning2020,zaverkinUncertaintybiasedMolecularDynamics2024}, gradually extends datasets with informative structures, resulting in more compact datasets that cover large parts of the relevant configurational space.

Dataset creation was further improved by the use of enhanced sampling methods, which increase structural diversity in the collected configurations.
The initial work by \citeauthor{herrMetadynamicsTrainingNeural2018a}\cite{herrMetadynamicsTrainingNeural2018a} utilized the root mean square deviation (RMSD) between the current configuration and a series of previous ones as a collective variable.
A metadynamics-like bias potential\cite{laioEscapingFreeenergyMinima2002} was applied to promote the exploration of diverse regions within configurational space.
More recent approaches have been tailored more specifically for MLIPs.
\citeauthor{yooMetadynamicsSamplingAtomic2021} \cite{yooMetadynamicsSamplingAtomic2021} have used the descriptor of high-dimensional neural network potentials as the collective variable instead of the RMSD.

Uncertainty-driven dynamics\cite{kulichenkoUncertaintydrivenDynamicsActive2023} and hyperactive learning\cite{novoselovMomentTensorPotentials2019,vanderoordHyperactiveLearningDatadriven2023,zaverkinUncertaintybiasedMolecularDynamics2024} use the model’s uncertainty to bias the system towards regions where the model is less confident.
Contour exploration\cite{watersEnergyContourExploration2021a,watersBenchmarkingStructuralEvolution2022a} evolves the system along constant potential energy contours.
The position updates are only limited by the local curvature of the PES thereby allowing for significantly larger position updates than MD, without relying on uncertainty estimates like uncertainty-driven dynamics.

Each of these provides a significant improvement in sampling efficiency over unbiased MD at a constant temperature.
However, these methods also have certain limitations.
The approach by \citeauthor{yooMetadynamicsSamplingAtomic2021},\cite{yooMetadynamicsSamplingAtomic2021} which performs metadynamics in descriptor space, constructs per-atom bias potentials, requiring a large number of descriptor comparisons and potentially leading to significant computational overhead.
Uncertainty-based methods, while effective in driving exploration toward regions of high model error, do not explicitly account for the separation in timescales of different degrees of freedom.
Intermolecular forces, such as those determining dynamical observables in liquids, are significantly smaller than intramolecular ones.
Thus, if the target quantity is small and underestimated, a calibrated uncertainty estimate will also be small, and uncertainty-based methods will not significantly enhance sampling along slow degrees of freedom.
This highlights a core conceptual difference between sampling objectives: increasing epistemic uncertainty versus increasing input diversity.
Uncertainty-driven approaches are reactive; they rely on the model effectively identifying its own knowledge gaps, leaving them vulnerable to poor calibration or noise.
Conversely, input diversity-driven approaches aim to maximize the volume of explored descriptor space independent of model error.
By forcing the system to populate underrepresented regions of the descriptor manifold, they ensure robust generalization and comprehensive phase space coverage.

In this work, we introduce Enhanced Representation-Based Sampling (ERBS), a novel enhanced sampling method for efficiently generating training data for MLIPs.
Starting from the mean descriptor of the system, we extract a small set of collective variables (CVs) via principal component analysis (PCA)\cite{stewartEarlyHistorySingular1993}.
Using these CVs, we construct a bias potential based on the recently introduced OPES-Explore framework\cite{invernizziExplorationVsConvergence2022}.
The combination of these CVs and bias potential allows for a rapid exploration of configurational space by following trajectories that sample preferentially along the $k$ maximum variance components of the MLIP features.

This work is structured as follows.
We begin with a description of the ERBS method.
Its usefulness in creating a static, non-active learned dataset is explored for the alanine dipeptide system.
Here, a screening of the bias parameters is performed, and the resulting configurational space coverage is analysed.
For some sets of these parameters, the generated trajectory is used as training data for MLIPs.
We first cross-validate the prediction metrics of models using both biased and unbiased validation datasets. Models trained on low- and high-temperature MD and ERBS trajectories are then used to compute the free energy surface (FES) of the dihedral angles in alanine dipeptide.
We find that the ERBS-trained models achieve lower errors with respect to the true FES compared to the MD-trained models, with the low-temperature MD model not producing a stable trajectory at all.

As a demonstration of ERBS use in an active learning setting, we turn to liquid water.
Two active learning workflows are set up, one using unbiased MD and one using ERBS biasing for sampling candidate configurations.
At each iteration of the workflow, the prediction metrics on the water dataset by \citeauthor{chengInitioThermodynamicsLiquid2019}\cite{chengInitioThermodynamicsLiquid2019} are calculated, and the diffusion coefficients are simulated.
We find that the prediction error on the literature test set decreases and diffusion coefficients converge to the value obtained from a model trained on the dataset by \citeauthor{chengInitioThermodynamicsLiquid2019} significantly faster for the ERBS run.

Finally, we compare the sampling behaviour of ERBS with that of UDD for the viscous room temperature ionic liquid 1-butyl-3-methylimidazolium tetrafluoroborate (\ce{BMIM+BF4-}).
Across a wide range of parameters, ERBS increases the mean squared displacement of up to 4 times compared to MD and 2 times compared to the best UDD result, indicating enhanced exploration of configurational space.

\section{Methods}

\subsection{Gaussian Moment Neural Network}

Our enhanced-sampling approach descriptor agnostic.
However, here we base it on the Gaussian-Moment Neural Network (GMNN)\cite{zaverkinGaussianMomentsPhysically2020,zaverkinFastSampleEfficientInteratomic2021} approach as it offers very fast training and inference times, while still achieving good prediction accuracy. Thus, GMNN is discussed here.

Given an atomic configuration $S$ consisting of Cartesian coordinates $\mathbf{R}$ and atomic numbers $Z$, potentials used in molecular dynamics map from $S$ to a potential energy $E$.
Most MLIPs utilize an atomic energy decomposition to predict energies for each local atomic environment\cite{behlerGeneralizedNeuralNetworkRepresentation2007a,unkeMachineLearningForce2021}.

\begin{align}\label{eq:energy_sum}
    E(S, \boldsymbol{\theta}) = \sum_i^{N_{\text{atoms}}} E_i( \boldsymbol{G}_i, \boldsymbol{\theta})
\end{align}

Restricting the range of interactions can be motivated by the short-sightedness of electronic matter\cite{Prodan_2005}, and the resulting linear scaling with system size has significantly contributed to these models' scalability.
The GMNN model consists of a descriptor, which constructs an invariant representation of each atom, and neural networks for predicting atomic energies.

First, the pairwise distances between a central atom i and its neighbors j are expanded in a radial basis, using Gaussian\cite{schuttSchNetDeepLearning2018a} or Bessel functions\cite{kocerNovelApproachDescribe2019}.
Embedding parameters $\beta_{Z_i, Z_j, n', n}$ are used to form linear combinations of the $n'$ original basis functions dependent on the atomic numbers of the central and neighboring atoms, $Z_i$ and $Z_j$.
This results in a contracted radial channel n.
Finally, angular information is captured by Cartesian moments, i.e., polynomials of the unit distance vectors $\boldsymbol{\hat{r}}_{ij}$ up to some rotation order $L$.

\begin{align}\label{eq:basis_fn}
    \Psi_{i, L, n} = \sum_{j \neq i} R_{Z_i, Z_j, n}(r_{ij}, \beta_{Z_i, Z_j, n', n}) \boldsymbol{\hat{r}}_{ij}^{\otimes L}
\end{align}

The invariant descriptor $\boldsymbol{G}$ is obtained from fully contracting the equivariant features $\Psi_{i, L, n}$ according to
\begin{align}\label{eq:contraction}
    G_{i, n_1, n_2} &= (\Psi_{i, 1, n_1})_a (\Psi_{i, 1, n_2})_a  \nonumber \\ 
    &\;\; \vdots \\
    G_{i, n_1, n_2, n_3} &= (\Psi_{i, 1, n_1})_a (\Psi_{i, 3, n_2})_{a,b,c} (\Psi_{i, 2, n_3})_{b,c}. \nonumber
\end{align}
Atomic energies are predicted from neural networks as $E_i = \mathrm{NN}(\boldsymbol{G}_i)$ and adjusted by element-specific scaling and shifting parameters, $\sigma_{Z_i}$ and $\mu_{Z_i}$.
\begin{align}\label{eq:scale_shift}
    E_i = \sigma_{Z_i} \cdot \text{NN}(\boldsymbol{G}_i) + \mu_{Z_i}
\end{align}
Finally, the atomic energies $E_i$ are summed up as in \cref{eq:energy_sum}.
Forces are calculated as the gradient of the total energy using automatic differentiation.
All learnable parameters of the model are optimized using stochastic gradient-based optimization.
The loss function minimized during training contains terms for energy and force errors:
\begin{gather}\label{eq:loss_fn}
    \mathcal{L}(\boldsymbol{\theta}) =
    \sum_{k=1}^{N_{\text{train}}} \Biggl[
    \lambda_E ||E^{\text{ref}}_k - E(S_k, \boldsymbol{\theta})||^2_2 \nonumber \\
    + \lambda_F \sum_i^{ N_{ \text{atoms} }^{(k)} } \frac{1}{3 N_{\text{atoms}}^{(k)} } ||\boldsymbol{F}^{\text{ref}}_{i,k} - \boldsymbol{F}_i(S_k, \boldsymbol{\theta})||^2_2
    \Biggr].
\end{gather}
Here, $\lambda_E$ and $\lambda_F$ denote hyperparameters for weighting the energy and force loss contributions, respectively.

\subsection{Active Learning}

In cases where no previously existing dataset can be used for training an MLIP, a new one has to be created from scratch.
However, ab initio molecular dynamics simulations are an expensive way to create training data as subsequent time steps are highly correlated.
Consequently, a common approach is to start from a handful of samples, train an initial model, and use it to sample new candidate structures.
A crucial step during sampling simulations is to terminate the trajectories when the model predictions become too inaccurate.
Estimating the error in model predictions is known as uncertainty quantification and is an active area of research\cite{pernotLongRoadCalibrated2022,bigiPredictionRigidityFormalism2024,tanEnhancedSamplingRobust2025,buskGraphNeuralNetwork2023a,heidSpatiallyResolvedUncertainties2024}.

In the present work, we use shallow ensembles recently proposed by \citeauthor{kellnerUncertaintyQuantificationDirect2024a}\cite{kellnerUncertaintyQuantificationDirect2024a}.
A shallow ensemble shares the weights for all but the last linear layer.
Instead of predicting a single atomic energy as in \cref{eq:scale_shift}, the model instead predicts $N_{\text{ens}}$ values, and the uncertainty of a prediction can be estimated from the sample standard deviation of the ensemble.
\begin{equation}\label{eq:uncertainty}
    \sigma_x  = \sqrt{ \frac{1}{N_{\text{ens}}} \sum_m^{N_{\text{ens}}} (x^{(m)} - \bar{x})^2 } 
\end{equation}
The mean of the ensemble predictions is used to drive the dynamics.

A crucial aspect is the calibration of the predicted uncertainty, i.e., how well predicted uncertainty and true error correlate. By training a shallow ensemble on a probabilistic loss function, like the negative log likelihood (NLL), miscalibrated uncertainty estimates are directly penalized during training.
\begin{align}
    \mathrm{NLL} = \frac{1}{2} \left[ \frac{(x - x_\text{ref})^2}{\sigma^2} + \log 2 \pi \sigma^2 \right]
    \label{eq:nll}
\end{align}

Once a sampling simulation has completed or was terminated after exceeding an uncertainty threshold, the trajectory represents a pool of candidate data and new datapoints can be selected from it.
Given a pool $\mathcal{D}_\text{pool} = {S_1, ..., S_n}$, batch active learning methods select a subset of the structures $\mathcal{D}_\text{batch} \subset \mathcal{D}_\text{pool}$ that maximizes an acquisition function $a$\cite{kirschBatchBALDEfficientDiverse2019a}, which may depend on the model parameters:
\begin{align}\label{eq:bal}
    \mathcal{D}_\text{batch} = \argmaxA_{\{S_1, ... , S_b\} \subset \mathcal{D}_\text{pool}} a(\{S_1, ... , S_b\}, \boldsymbol{\theta})
\end{align}

Throughout this work, we use greedy maximum-distance selection with a last-layer gradient feature map\cite{zaverkinExploringChemicalConformational2022}:
\begin{align}
    \phi_{\text{ll}}(S) &= \nabla_{\theta_{\text{ll}}} E(S, \theta) \label{eq:lastlayer}\\
    S &= \argmaxA_{S \in \mathcal{D}_\text{pool} / \mathcal{D}_\text{batch}} \min_{S' \in \mathcal{D}_\text{pool} \cup \mathcal{D}_\text{batch}} || \phi_{\text{ll}}(S) - \phi_{\text{ll}}(S') ||_2\label{eq:maxdist}
\end{align}
The feature map $\phi_{\text{ll}}(S)$ serves to compare the similarity of two structures in terms of the model's last layer weight gradient.
\Cref{eq:maxdist} is applied iteratively to select structurally diverse points in feature space, a selection algorithm also known as farthest point sampling\cite{gonzalezClusteringMinimizeMaximum1985}.
More details on the selection methods can be found in the original work by \citeauthor{zaverkinExploringChemicalConformational2022}\cite{zaverkinExploringChemicalConformational2022}

\begin{figure}
 \centering
 \includegraphics[width=8cm]{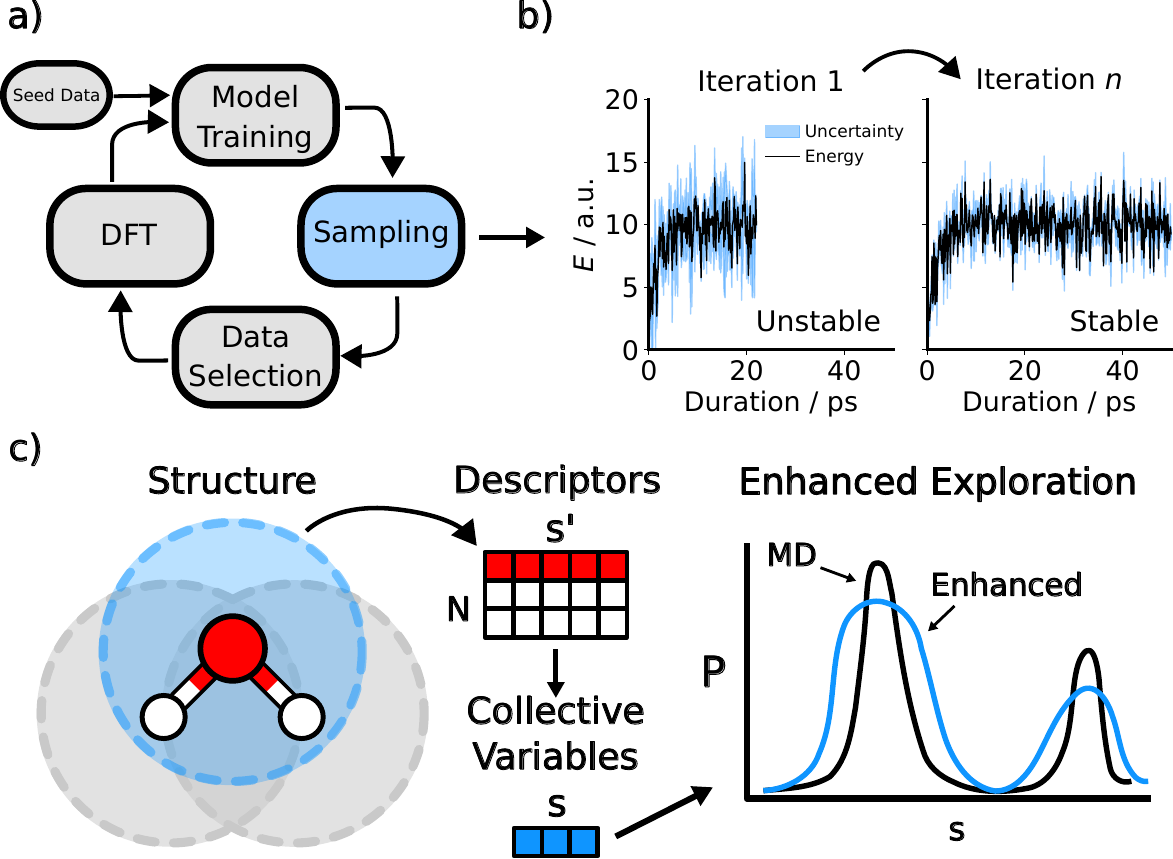} 
 \caption{
    (a) Illustration of an active learning cycle: Starting from seed data, models are iteratively trained and used to sample candidate configurations. From these candidates, the most informative ones are selected for DFT calculations.
    (b) Sampling stability improves and predicted uncertainties decrease over active learning iterations.
    (c) Workflow for constructing the ERBS potential: high-dimensional atomic descriptors $\mathbf{s'}$ are aggregated and reduced to collective variables $\mathbf{s}$ via dimensionality reduction.
    The bias potential flattens the sampled distribution $P$.
 }
 \label{fig:active-learning}
\end{figure}

\Cref{fig:active-learning}a) depicts the typical steps involved in an active learning loop.
As training iterations proceed, model uncertainty decreases, and the quality and stability of simulated trajectories increase (\cref{fig:active-learning}b). 
Viewed differently, it takes increasingly longer for the model to encounter informative new configurations during a sampling simulation.
As a result, the sampling trajectories need to become progressively longer.

In order to increase the diversity of sampled structures, enhanced sampling methods for the generation of MLIP training data have been developed.
One such approach, UDD, builds directly on the model's uncertainty estimate and uses it to steer the dynamics towards more uncertain regions.
In UDD, a bias potential is constructed to encourage exploration of regions with high uncertainty\cite{kulichenkoUncertaintydrivenDynamicsActive2023}:
\begin{align}
    E_{\text{UDD}}(\sigma_E^2) = A \left[ \exp\left( -\frac{\sigma_E^2}{N_{\text{ens}} N_{\text{atoms}} B^2} \right) - 1 \right]
    \label{eq:udd}
\end{align}
Here, $A$ and $B$ are empirically determined parameters determining the strength of the bias and its gradient.
The MD is then propagated by the sum of the MLIP prediction and $E_{\text{UDD}}$.
The methods described so far and the GMNN architecture are implemented in \texttt{apax}\cite{schaferApaxFlexiblePerformant2025}, which is used for all model trainings and MD simulations in this work.

\subsection{Enhanced Representation Based Sampling}

Improving configurational space sampling for MLIP data generation can be framed as exploring diverse inputs to the model.
To an MLIP like GMNN, the descriptor represents the model's input and provides a general-purpose set of collective variables for identifying undersampled regions in configuration space.
For the present method, we use the average descriptor vector of the entire system.
\begin{equation}\label{eq:meandescriptor}
    \mathbf{s}' = \frac{1}{N_{\text{atoms}}}\sum_i^{N_{\text{atoms}}} \boldsymbol{G}_i
\end{equation}

Constructing a global descriptor ensures differentiability and computational efficiency and has found various applications in dataset analysis and data selection\cite{deComparingMoleculesSolids2016b,bartokMachineLearningUnifies2017,roweAccurateTransferableMachine2020a}.
Here, we use a variant of the descriptor in \cref{eq:contraction} without element-dependent parameters, $\boldsymbol{\beta}$. However, including them or using other model features would also be possible.
Over the course of a simulation, system-averaged descriptors are collected at fixed time intervals and used as reference descriptors to compare the similarity between the current configuration and past ones.

The descriptor is high-dimensional, and its entries are correlated with each other.
As a result, sampling the descriptor space directly is challenging due to the curse of dimensionality.
A reduced set of CVs is obtained from principal component analysis (PCA)\cite{stewartEarlyHistorySingular1993}, which identifies the most relevant collective motions in descriptor space.
PCA is a common dimensionality reduction method that has also seen various applications in CV-based enhanced sampling\cite{chenCollectiveVariableDiscovery2018,sicardReconstructingFreeenergyLandscape2013}.
As usual for PCA, the data matrix is first centred using the per-feature mean $\boldsymbol{\mu}$.

\begin{equation}
    \mathbf{\hat{S}} = \mathbf{S}' - \mathbf{1}_{N_{\text{ref}}}\boldsymbol{\mu}^\text{T}
\end{equation}

Here, the data matrix $\mathbf{S}'$ consists of stacked reference descriptors $\mathbf{s}'$, and $\mathbf{\hat{S}}$ represents its centered version.
The principal components $\mathbf{V}$ are obtained from a singular value decomposition of $\mathbf{\hat{S}}$.
By truncating $\mathbf{V}$ to the first $k$ columns, we obtain a projection matrix, $\mathbf{V}^{(k)}$,  used to reduce the dimensionality of the descriptor.
During sampling simulations, the CVs are computed via the feature dimensionality reduction function $\phi$.

\begin{equation}\label{eq:dimred}
    \phi(\mathbf{s}') = (\mathbf{s}' - \boldsymbol{\mu}) \mathbf{V}^{(k)} = \mathbf{s}
\end{equation}

Although PCA is used throughout all experiments, we emphasize that other choices of $\phi$, such as autoencoders\cite{kingmaAutoEncodingVariationalBayes2022}, are possible.
Based on these CVs, a bias potential is constructed similarly to the ``explore'' variant of On-the-Fly Probability Enhanced Sampling (OPES)\cite{invernizziRethinkingMetadynamicsBias2020,invernizziExplorationVsConvergence2022}.
While OPES estimates the unbiased probability distribution, OPES-explore estimates the well-tempered one.
Since the well-tempered distribution is smoothed out, fewer kernels are required to estimate it\cite{barducciWellTemperedMetadynamicsSmoothly2008}.
The probability density of the CV space is modelled on-the-fly by depositing Gaussian kernels, $K$, at fixed intervals during a molecular dynamics simulation centered on the current averaged descriptor $\mathbf{s}_{j}$.

\begin{equation}\label{eq:kernel}
    K(\mathbf{s}, \mathbf{s}_{j}) = \frac{1}{\sqrt{ \det{(\boldsymbol{\Sigma})} (2\pi)^k}} \exp \left( - \frac{1}{2} (\mathbf{s} - \mathbf{s}_j)^\text{T} \boldsymbol{\Sigma}^{-1} (\mathbf{s} - \mathbf{s}_j) \right)
\end{equation}

We restrict ourselves to isotropic covariances $\boldsymbol{\Sigma}$ for computational efficiency. 
As we implement the kernel compression algorithm from \citeauthor{invernizziRethinkingMetadynamicsBias2020}\cite{invernizziRethinkingMetadynamicsBias2020}, these may become diagonal over the course of a simulation.

The overall well-tempered probability density $p_n^\text{WT}(\mathbf{s})$ is, thus, modelled by the average over all kernels $K(\mathbf{s}, \mathbf{s}_{j})$.
\begin{equation}\label{eq:probability}
    p_n^\text{WT}(\mathbf{s}) = \frac{1}{N_{\text{ref}}}\sum_j^{N_{\text{ref}}} K(\mathbf{s}, \mathbf{s}_j)
\end{equation}
The OPES-explore bias potential at time step $n$ can then be calculated from the probability density as 
\begin{equation}\label{eq:biaspot}
    V_n(\mathbf{s}) = (\gamma - 1) \frac{1}{\beta} \log \left( \frac{p_n^\text{WT}(\mathbf{s})}{Z_n} + \epsilon\right)    
\end{equation}
$Z_n$ is a modified normalisation constant computed from numerically integrating the density with the kernel centers as integration points. The parameter $\epsilon$ contains a barrier parameter $\Delta E$ via $\gamma = \beta \Delta E$, where $\beta=1/(k_{\text{B}} T)$ is the inverse thermal energy.
\begin{equation}\label{eq:barrier}
    \epsilon = \exp \left( \frac{- \gamma}{\gamma - 1} \right)
\end{equation}
The incorporation of $\Delta E$ allows the method to place a soft limit on the maximum strength of the bias potential.
OPES-explore offers several advantages over metadynamics for configurational space exploration.
While metadynamics and its variants slowly deposit Gaussian bias hills, OPES-explore models the probability density.
As $p_n^\text{WT}$ is normalized, the simulation starts with a strong bias right from the start.
Further, the normalization constant is modified in such a way that it only increases when kernels overlap, improving exploration.
A more detailed discussion of these advantages and the construction of $Z_n$ can be found in the original publication by \citeauthor{invernizziRethinkingMetadynamicsBias2020}\cite{invernizziRethinkingMetadynamicsBias2020}.
The conceptual steps involved in the ERBS method are illustrated in \cref{fig:active-learning}.
It should be noted that a simulation can either use a fixed PCA basis constructed from a preexisting dataset or use a variable one that is recomputed every time a new reference descriptor is added.
For a fixed dimensionality reduction function, the target-probability density in \cref{eq:probability} is well-defined.
As the goal of ERBS is not to be a free energy method, but to sample the most structurally diverse configurations within a physically meaningful energy range, we use a variable basis for all experiments.

Finally, to analyse the scalability of ERBS, we consider the computational cost associated with the bias force evaluation.
The gradient of the bias potential in \cref{eq:biaspot} can be written as

\begin{equation}
    \frac{\partial V_n}{\partial \mathbf{R}} = \frac{\partial V_n}{\partial \mathbf{s}} \cdot \frac{\partial \mathbf{s}}{\partial \mathbf{R}}.
\end{equation} \label{eq:chainrule}

Here, $\frac{\partial \mathbf{s}}{\partial \mathbf{R}}$ is the Jacobian of the reduced descriptor vector $s$ with respect to the atomic positions $R$.
Notably, it is independent of the number of reference descriptors.
The term $\frac{\partial V_{n}}{\partial \mathbf{s}}$, on the other hand, involves the sum of kernel gradient contributions from each of the reference descriptors and thus scales linearly with the number of references.
However, since the first term consists of simple algebraic operations, its cost is negligible compared to that of evaluating $\frac{\partial \mathbf{s}}{\partial \mathbf{R}}$.

As a result, the overall cost of ERBS using the GM descriptor is comparable to the cost of a GMNN force evaluation and remains practically independent of the number of reference descriptors used for even extensive active learning sampling runs.
A comparison of the scaling behaviour for ERBS and the method by \citeauthor{yooMetadynamicsSamplingAtomic2021}\cite{yooMetadynamicsSamplingAtomic2021} can be found in the SI.

\section{Results}

\subsection{Static dataset generation for Alanine Dipeptide}

Alanine dipeptide is a commonly used test system in the enhanced-sampling literature\cite{laioMetadynamicsMethodSimulate2008,tanEnhancedSamplingRobust2025} due to the thoroughly investigated FES in two of its dihedral angles, $\Phi$ and $\Psi$.
Consequently, the degree to which a trajectory covers the dihedral angle space serves as an indication of how well a general-purpose sampling method is able to identify physically relevant degrees of freedom.
In order to examine the sensitivity of ERBS to the particular choice of parameters, we perform an extensive parameter scan on alanine dipeptide in vacuum.
The scanning ranges for the parameters in \cref{eq:dimred,eq:kernel,eq:barrier} are selected as follows.
For the barrier parameter in \cref{eq:barrier}, we use $\Delta E = \{ 5, 10, 15, 20, 25\}$ eV. The covariance matrix in \cref{eq:kernel} is chosen as $\Sigma=\sigma^2 \mathbf{I}_k$, with $ \sigma = \{ 0.05, 0.1, 0.2, 0.5\}$, corresponding to isotropic kernels in a reduced feature space.
Finally, the dimensionality of the descriptor space in \cref{eq:dimred} is selected from $k = \{ 2, 4, 6, 8, 10\}$.

\begin{figure}
 \centering
 \includegraphics[width=8cm]{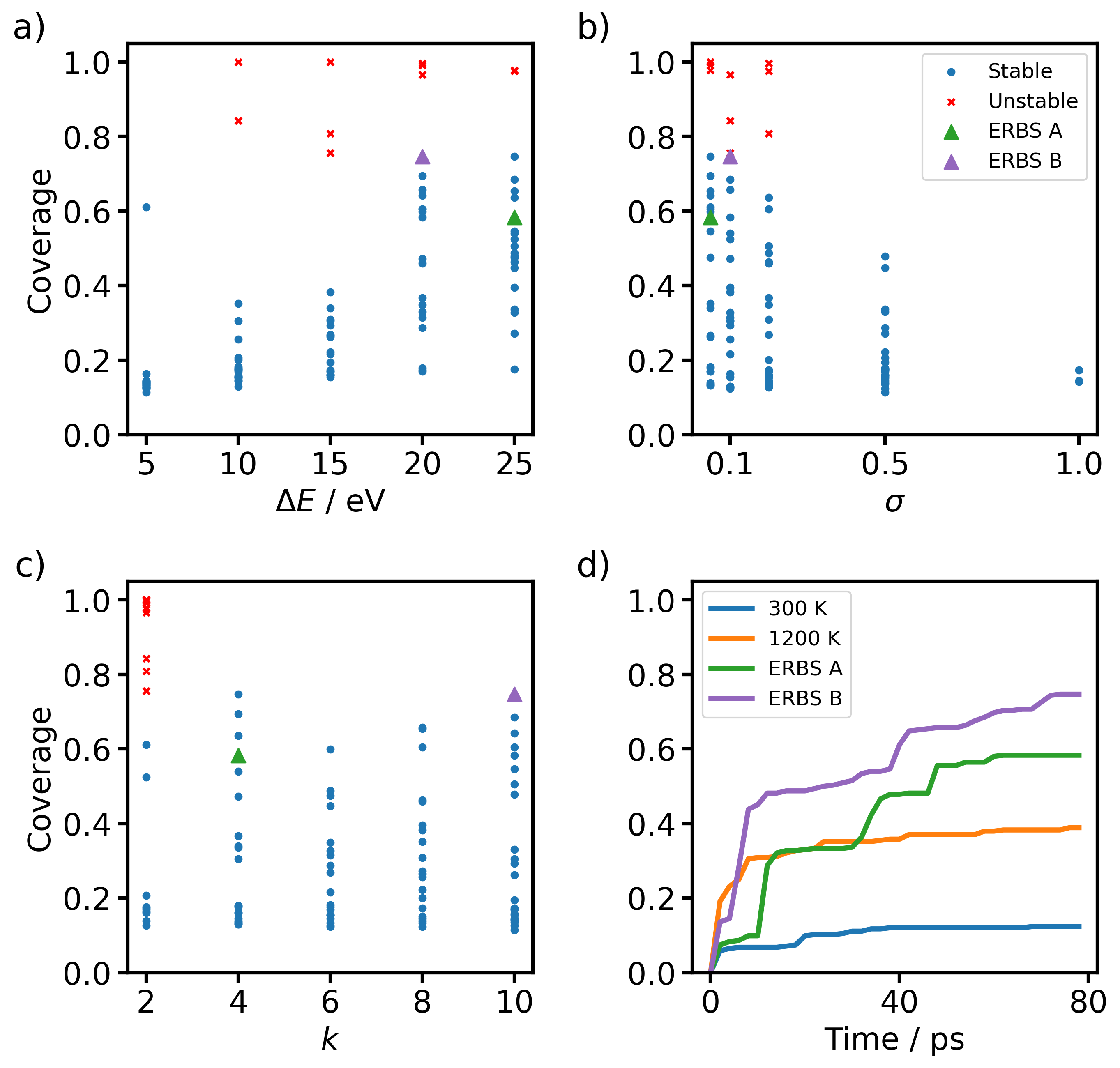} 
 \caption{
    a--c) Coverage vs parameter choices, d) $\Phi-\Psi$-space coverage of alanine dipeptide over time using molecular dynamics and the enhanced sampling method. 
 }
 \label{fig:exploration}
\end{figure}

For each possible parameter combination, an \SI{80}{\pico\second} trajectory is simulated using a Langevin thermostat at \SI{300}{\kelvin} with a timestep of \SI{0.5}{\femto\second} using the Amber99-SB force ﬁeld\cite{wuSignificantlyImprovedProtein2017,guoProteinAllosteryConformational2016} in CP2K\cite{kuhneCP2KElectronicStructure2020}.
New kernels were deposited every 10,000 steps.
In addition to the biased simulations, we also compute unbiased trajectories at \SI{300}{\kelvin} and \SI{1200}{\kelvin} to set a baseline for the system's exploration and to verify whether the effect of the bias could also be reached by increasing the temperature.
The $\Phi-\Psi$-space coverage for the unbiased and biased trajectories is displayed in \cref{fig:exploration}.
In order to track and quantify the coverage, the space was tiled into 15°-by-15° squares.
Coverage is then calculated by the ratio of visited squares to their total amount.

The initial structure is located in a free energy minimum. The unbiased dynamics are stuck there for the entire duration of the trajectory; hence, the coverage is severely limited.
A few parameter choices with $k=2$ lead to the dissociation of the molecule.
This is expected for high barrier heights and small Gaussian bandwidths, which lead to such strong forces that the simulation becomes unstable.
While the number of physically relevant collective variables in this system is two, ERBS learns them on the fly and, thus, needs time to identify them over the course of a simulation.
It is thus advisable to initially overestimate the value of $k$, since there is only a negligible cost associated with an increase in $k$ we largely find an insensitivity of the coverage to the particular choice of the number of principal components.
The lack of a trend for $k\geq2$ demonstrates that the PCA successfully concentrated the relevant slow dynamics into the first few principal components.
Otherwise, the exploration was drastically increased when compared to the unbiased simulation at \SI{300}{\kelvin}, and most parameter choices outperformed even the \SI{1200}{\kelvin} trajectories exploration with coverages of up to \SI{75}{\percent}.
The parameter set leading to the largest coverage used $\Delta E = 20$ eV,  $\sigma=0.05$ and $k=4$
We will refer to the trajectory with the best coverage as ERBS B.
We also highlight a second trajectory which also achieves good coverage but with a completely different set of parameters, $\Delta E = 25$ eV,  $\sigma=0.1$ and $k=10$ and refer to it as ERBS A.

Next, the suitability of the biased trajectories as MLIP training data was investigated. 
Four trajectories were chosen for dataset creation: the unbiased MD trajectories at \SI{300}{\kelvin} and \SI{1200}{\kelvin} and the biased trajectories ERBS A and ERBS B
The datasets contain 1800 training and 200 validation samples, respectively.
In each case, the data points are selected randomly, but the validation samples are taken from the last 20 percent of the trajectory.
GMNN models were trained on each dataset using identical hyperparameters.
A radial basis consisting of 16 spherical Bessel functions was chosen with a cut-off of \SI{5}{\angstrom}.
Two neural network layers of size 64 were used, and the model was trained with the AdamW optimizer\cite{DBLP:journals/corr/KingmaB14,loshchilovDecoupledWeightDecay2019}. 
Further training details can be found in \cref{tab:si:hypers}.

\begin{figure}
 \centering
 \includegraphics[width=8cm]{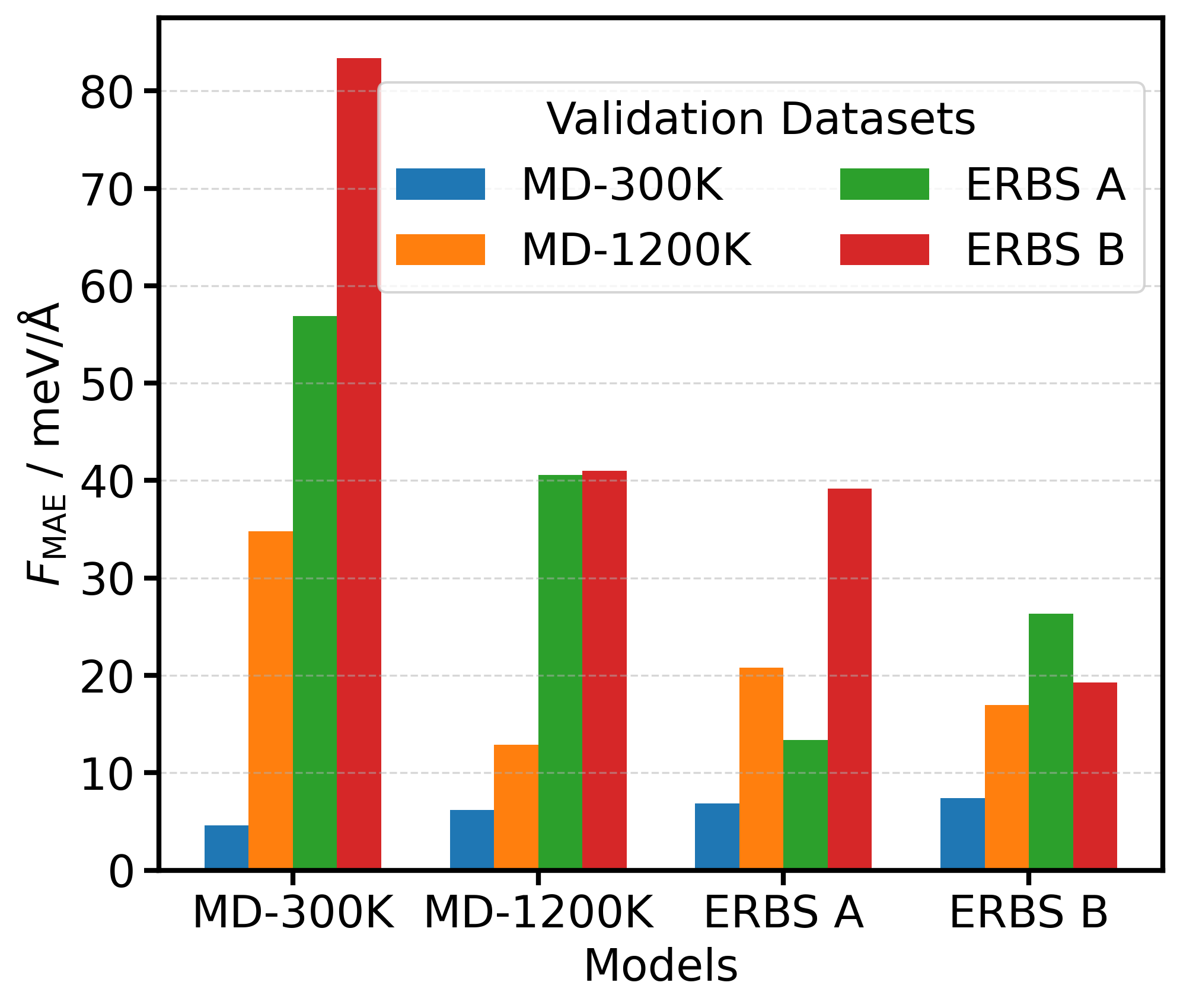} 
 \caption{
    Cross-validation of force MAEs.
    The x-axis indicates the training dataset used to generate each model.
    The colored bars within each group represent the model's performance on the four distinct validation datasets, as defined in the legend.
 }
 \label{fig:aladmetrics}
\end{figure}

The force MAEs of each model evaluated on each dataset are displayed in \cref{fig:aladmetrics}.
\begin{equation}
    F_\mathrm{MAE} = \frac{1}{N_{\text{atoms}} N_{\text{structures}}} \sum_{j=1}^{N_{\text{structures}}} \sum_{i=1}^{N_{\text{atoms}}} \left| \mathbf{F}_i^{\mathrm{pred}, (j)} - \mathbf{F}_i^{\mathrm{ref}, (j)} \right|
\end{equation}

For each validation dataset, the model trained on the corresponding training dataset achieves the lowest error.
While the model trained directly on the \SI{300}{\kelvin} data achieves the lowest errors on the \SI{300}{\kelvin} validation set, it performs the worst on all other validation sets.
Conversely, the model trained on the \SI{1200}{\kelvin} MD data generalizes better to other MD-derived datasets but ranks near the bottom on the ERBS validation sets.
In contrast, models trained on ERBS-sampled data perform consistently well across biased and unbiased validation sets, demonstrating strong transferability.
More detailed prediction-error parity plots for the four models can be found in \cref{fig:si:parity_alad}.

\begin{figure}
 \centering
 \includegraphics[width=8cm]{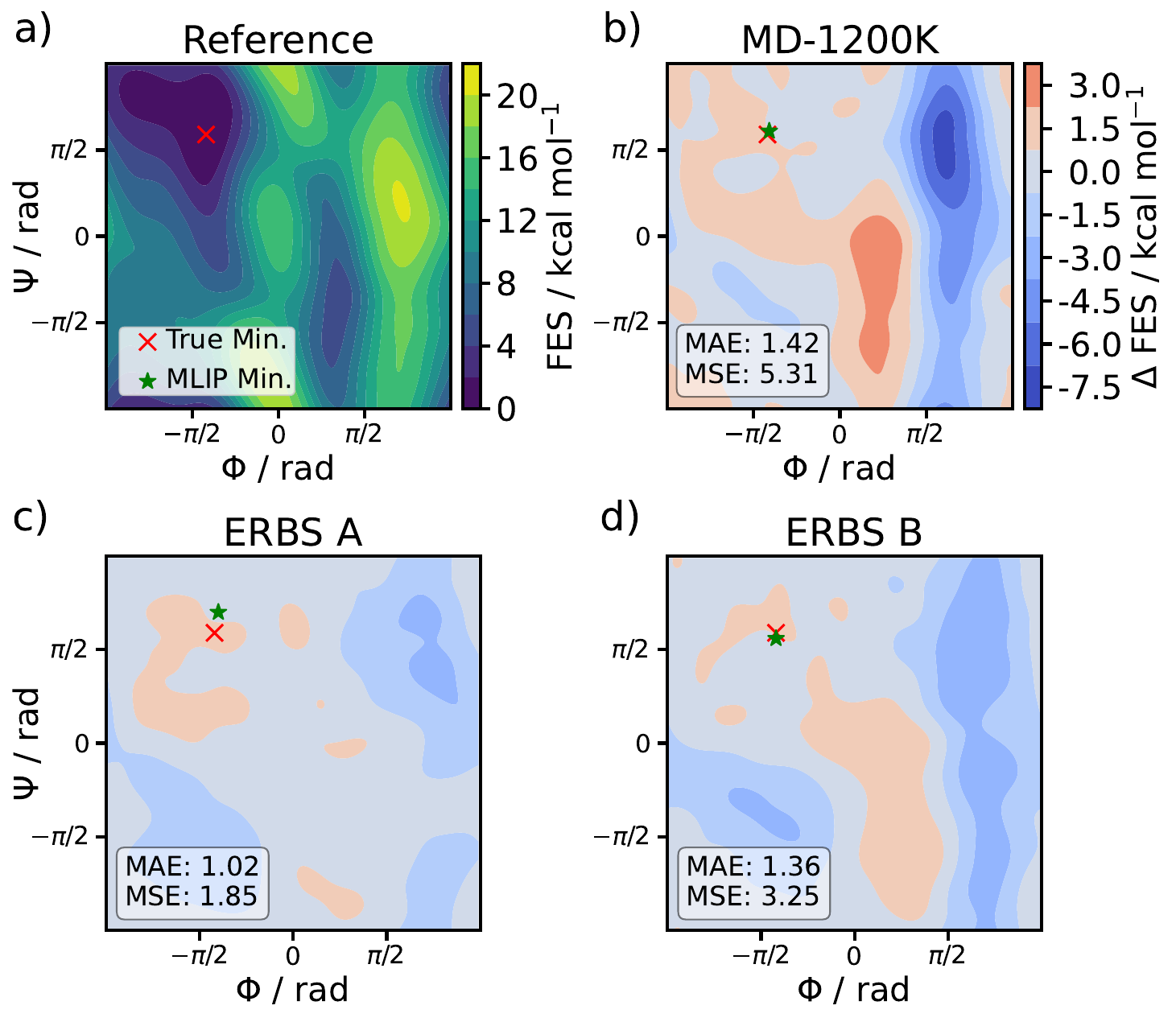} 
 \caption{
    (a) Free energy surface (FES) of alanine dipeptide computed with the reference force field.  
    (b–d) Signed errors ($\Delta$FES = FES$_{\text{MLIP}}$ - FES$_{\text{Reference}}$) between the reference and the FES computed with models trained on data from (b) MD-1200 K, (c) ERBS A, and (d) ERBS B.  
    Blue regions indicate lower predicted free energies relative to the reference, while red regions indicate overestimation.
    The red cross and green stars indicate the location of the Global free energy minimum of the reference force field and trained MLIPs, respectively.
 }
 \label{fig:fescomp}
\end{figure}

While validation metrics give a reasonable first estimate of model performance, the purpose of MLIPs is to compute physical observables, such as FESs.
The FES in $\Phi$ and $\Psi$ can be evaluated using models trained on configurations from biased and unbiased dynamics.
Well-tempered metadynamics simulations using Plumed\cite{tribelloPLUMED2New2014,bonomiPromotingTransparencyReproducibility2019} were conducted for the MD-300K, MD-1200K, and ERBS models trained on the datasets from the previous experiment.
A biasing factor of 10, hill size of \SI{1.2}{\kilo\joule\per\mole}, and bandwidth of 0.35~rad were used.
It should be emphasized that the parameters for metadynamics fulfil different purposes and can't be directly compared to the ones used in ERBS despite similar names.
Further, a ground truth simulation was carried out with the Amber99-SB force field.
All simulations were conducted in the NVT ensemble at \SI{300}{\kelvin} and lasted for \SI{10}{\nano\second}.
The reference free-energy surface and the errors of the FES produced by the models MD-1200K and the ERBS models are displayed in \cref{fig:fescomp}.
All FES were shifted such that the global minimum of each surface is set to zero.
This alignment allows for consistent comparison of relative free energy differences, and the error was computed with respect to the reference FES based on the aligned values.

The model trained on the \SI{300}{\kelvin} trajectory data is unstable. The system forms new bonds during the trajectory, which keeps it stuck in a non-physical free-energy minimum.
Despite the different parameter choices for the bias and resulting differences in the $\Phi$-$\Psi$ space coverage of models ERBS A and ERBS B, they achieve similar quality in their FESs.
While the MD-1200K model also achieves good metrics, it is surpassed by both models trained on enhanced sampling data.
In terms of localization, both the ERBS B and MD 1200 K models correctly identify the global free energy minimum, whereas ERBS A exhibits a minor deviation.
However, the MD 1200 K model yields significantly larger errors in the broader free energy surface.
This discrepancy can be attributed to the training data: the 1200 K trajectory remained trapped within the global minimum basin, failing to explore the right-hand side of Ramachandran space.
The coverage of the three sampling trajectories is shown in \cref{fig:si:coverage_si}.
The best model, ERBS A, achieves a free energy MAE of \SI{1.02}{\kilo\calorie\per\mole}, which is almost chemically accurate, but could easily be refined for production accuracy within a few active learning iterations.

In a recent study by \citeauthor{tanEnhancedSamplingRobust2025}\cite{tanEnhancedSamplingRobust2025}, the data efficiency of MLIPs in reconstructing the alanine dipeptide FES was investigated.
Using their novel eABF method, they perform active learning for the same system and conclude that an accurate FES may be accessible at 4000 data points.
The datasets generated using the ERBS method were taken from static \SI{80}{\pico\second} trajectories and are comprised of 2000 datapoints, suggesting that chemical accuracy could be achieved with significantly less than 4000 data points when using ERBS biasing in an active learning setting.

\subsection{Accelerated Active Learning of Liquid Water}

In order to demonstrate the efficacy of ERBS in an active learning setting, we consider the case of liquid water.
The initial system was constructed using Packmol\cite{martinezPACKMOLPackageBuilding2009} with 32 water molecules in a periodic box at the experimental density of \SI{997}{\kilo\gram\per\cubic\meter}.
Energy and forces of all data points collected in the active learning scheme were computed using the revPBE0 hybrid functional\cite{adamoReliableDensityFunctional1999} with a plane wave cutoff of 400 Ry, TZV2P-GTH basis sets\cite{goedeckerSeparableDualspaceGaussian1996}, Goedecker--Teter--Hutter pseudo potentials\cite{beckeDensityfunctionalThermochemistrySystematic1997a,hartwigsenRelativisticSeparableDualspace1998,krackPseudopotentialsKrOptimized2005} and the D3 dispersion correction\cite{grimmeConsistentAccurateInitio2010,grimmeEffectDampingFunction2011a} in CP2K\cite{kuhneCP2KElectronicStructure2020}.
All DFT parameters are adapted from \citeauthor{chengInitioThermodynamicsLiquid2019}\cite{chengInitioThermodynamicsLiquid2019}.

Starting from the Packmol structure, 200 configurations were bootstrapped by randomly rotating and translating molecules and atomic displacements.
The first 2 iterations of active learning were performed on these bootstrapped samples as follows.
A first model was trained on 10 and validated on 6 randomly selected data points.
From the remaining data pool, another 5 training samples were chosen using the maximum distance selection with last-layer gradient features.

Afterwards, two AL workflows were set up.
Both are identical apart from the use or lack of the ERBS bias potential.
For the biased workflow, a new kernel was placed every \SI{5}{\pico\second} with a barrier height of \SI{0.5}{\electronvolt} per atom and a bandwidth of 1.0.
The first 3 PCA components were used in the construction of the CVs.
GMNN models were trained as shallow ensembles with 16 members in order to have access to uncertainty estimates.
The radial cutoff was chosen to be \SI{5.5}{\angstrom} and a neural network with two hidden layers of 128 and 64 units was used throughout.
The remaining hyperparameters are listed in \cref{tab:si:hypers}.

For every iteration, a \SI{25}{\pico\second} sampling simulation was set up using a time step of \SI{0.5}{\femto\second} and a coupling constant of \SI{500}{\femto\second} for the Berendsen thermostat.
A force uncertainty threshold of \SI{3.0}{\milli\electronvolt\per\angstrom} was used as a stopping criterion for the trajectories.
The sampling simulations were conducted with the \texttt{ASESafeSampling} node in IPSuite.
Upon reaching the uncertainty threshold, the geometry is reset to the starting configuration, the momenta are initialized, and the simulation continues until the specified duration is reached.
After each sampling simulation, 4 data points were chosen randomly for the validation dataset, and 10 data points were chosen using maximum distance selection with last-layer gradient features from the trajectory.
For the remainder of the active learning cycles, these two methods were used for all training and validation data selections, respectively.
In total, 10 active learning iterations were conducted for both setups, resulting in 115 training and 46 validation samples in each case.
The prediction-error parity plots for the active learned models as well as the model trained on the literature dataset can be found in \cref{fig:si:parity_water}.

\begin{figure}
 \centering
 \includegraphics[width=8cm]{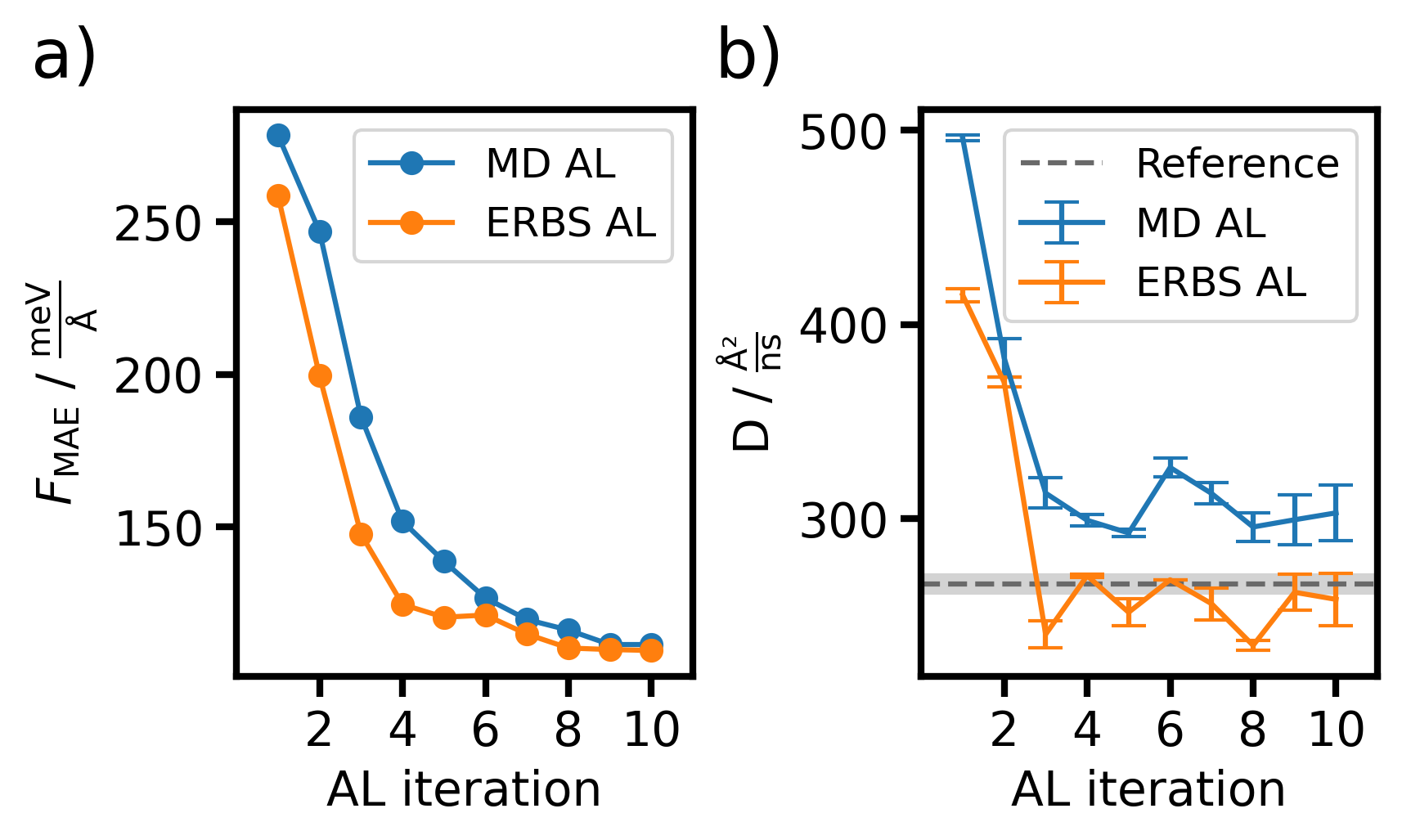} 
 \caption{
    a) Force mean absolute errors with respect to the literature water dataset
    across active learning iterations for the models created with MD and ERBS-based active learning. b) Diffusion coefficients of water simulated using models from successive active learning iterations, compared to the value obtained with the model trained on the literature dataset. 
 }
 \label{fig:water}
\end{figure}

The water dataset by \citeauthor{chengInitioThermodynamicsLiquid2019}\cite{chengInitioThermodynamicsLiquid2019} was generated with an emphasis on structural diversity and at in part sampled using path-integral molecular dynamics.
Models trained on it were shown to achieve good agreement with experiment for structural properties of liquid water, relative stabilities of different phases of water ice and other thermodynamic observables.
As we use the same level of theory and DFT code for labelling data in this experiment, we can evaluate all models on their dataset.
\Cref{fig:water} a) displays the Force MAE for each model iteration from the active learning runs with and without enhanced sampling compared to a model fitted directly to the literature training data.
We observe that the models trained using ERBS-enhanced sampling consistently achieve lower force errors over all AL iterations compared to those trained using standard MD AL sampling.
As the number of AL cycles increases, the performance gap between the two approaches narrows.
It is to be expected that neither workflow achieves chemical accuracy on the literature test dataset.
The test dataset includes configurations of varying densities and bond breakages, neither of which is contained in the datasets constructed here.

In addition to test set metrics, we simulate diffusion coefficients to investigate the ability of the two AL setups in producing models that can accurately simulate experimental observables.
We simulate \SI{5}{\nano\second} trajectories of 256 water molecules in the NVT ensemble at \SI{300}{\kelvin} using the models obtained from each AL iteration.
Additionally, the diffusion was also simulated with the model trained on the literature dataset in order to rule out model-architecture-specific limitations in describing the dynamics. 
The center-of-mass diffusion coefficients, $D$ are obtained from a linear fit to the mean-squared displacement (MSD)
\begin{equation}\label{eq:msd}
    \text{MSD}(\Delta t) = \left\langle \left| \frac{1}{N} \sum_{i=1}^{N} \mathbf{R}(t) - \mathbf{R}(0) \right|^2 \right\rangle = 6 D \Delta t
\end{equation}
Here, the sum goes over all $N$ equivalent particle positions, $\mathbf{R}$, or in this case the mass centers of the water molecules and $\Delta t$ is the simulation time.
Based on the diffusion coefficient thus calculated from a finite system, the infinite system-size limit is obtained via the Yeh–Hummer correction\cite{yehSystemSizeDependenceDiffusion2004}.

\begin{equation}\label{eq:diffusion}
D(\infty) = D(L) + \frac{\xi}{6 \pi \beta \eta L}
\end{equation}

The first term, $D(L)$ corresponds to the diffusion coefficient obtained from rearranging \ref{eq:msd} for a finite system of side length $L$.
The correction term includes the geometry-dependent constant $\xi = 2.837297$
(for a cubic box) and the shear viscosity $\eta$.

The first \SI{200}{\pico\second} of the trajectories are discarded for equilibration.
To estimate uncertainties, block averaging is applied.
All simulated diffusion coefficients and the experimental value are displayed in \cref{fig:water} b).

Diffusion coefficients simulated with the ERBS models are consistently closer the the reference value obtained with the model trained on the literature dataset across all AL iterations. 
Starting from iteration 4, the ERBS model diffusion coefficients reach good agreement with the value produced with the reference model, with the standard errors overlapping for 5 of the remaining 7 trajectories.
To further validate the active-learned models, we have computed the oxygen-oxygen radial distribution functions, which can be found in \cref{fig:si:water_rdf}.

While the reference and ERBS models yield similar diffusion coefficients, they both overestimate the experimental value of \SI{241}{\square\angstrom\per\nano\second}\cite{holzTemperaturedependentSelfdiffusionCoefficients2000} by about \SI{25}{\square\angstrom\per\nano\second}.
Possible reasons could either be limitations of the GMNN model or a tendency of the PBE0 hybrid functional to overestimate the diffusion.
Diffusion coefficients reported in AIMD studies strongly depend on the choice of density functional\cite{gillanPerspectiveHowGood2016,fernandez-serraNetworkEquilibrationFirstprinciples2004,leeDynamicalPropertiesLiquid2007}.
In contrast, \citeauthor{daruCoupledClusterMolecular2022} \cite{daruCoupledClusterMolecular2022} report excellent agreement between simulated and experimental diffusion coefficients using an MLIP trained on coupled cluster data, with simulations that also account for nuclear quantum effects.
It is worth noting that, due to the small datasets used here, adding a few data points can change the PES of the model quite drastically between iterations, and a monotonic convergence to the reference model cannot be expected.
Nevertheless, the rapid improvement and overall stability of the ERBS-trained models highlight the effectiveness of enhanced sampling in building accurate MLIPs with minimal data.

\subsection{Comparison to Uncertainty Based Sampling}

To benchmark the sampling performance of ERBS against UDD, we selected the ionic liquid \ce{BMIM+BF4-} as a test system.
Its inherently high viscosity characteristic poses a particular challenge for active learning, as long molecular dynamics trajectories are typically required to adequately explore intermolecular interactions\cite{salanneSimulationsRoomTemperature2015, bedrovMolecularDynamicsSimulations2019,shayestehpourEfficientMolecularDynamics2023c}.

A shallow ensemble is trained on a 200-structure subset of the \ce{BMIM+BF4-} dataset created by \citeauthor{zillsMachineLearningdrivenInvestigation2024}\cite{zillsMachineLearningdrivenInvestigation2024}.
The prediction-error parity plots for the model are shown in \cref{fig:si:parity_bmim}.
The model hyperparameters are identical to the experiment on liquid water from the previous section.
It achieves energy and force MAEs on the validation set of \SI{0.75}{\milli\electronvolt\per\atom} and \SI{62}{\milli\electronvolt\per\angstrom} respectively.
The quality of uncertainty estimates is discussed in \cref{calmetrics}.
Next, we perform sampling simulations with unbiased MD, UDD, and ERBS.
To ensure a fair comparison, we performed parameter scans for both UDD and ERBS.
We perform UDD for all combinations of $A = \{ 0.1, 0.5, 1.0, 2.0, 5.0, 10.0\}$~\si{\electronvolt\per\atom} and $B = \{0.1, 0.5, 1.0, 2.0 \}$~eV cf. \cref{eq:udd}.
As ERBS has more parameters, we focus on the barrier height, bandwidth, and number of principal components.
Starting from $\Delta E = 1.5$~\si{\electronvolt\per\atom}, $\sigma = 2.0$, and $k= 2$, one parameter at a time was scanned while holding the others fixed to reduce the number of simulations.
The individual parameter ranges were chosen as $\Delta E = \{ 0.1, 0.5, 1.5, 2.5\}$~\si{\electronvolt\per\atom}, $\sigma = \{ 0.5, 1.0, 2.0, 5.0\}$ and $k= \{ 1,2,3,4\}$.
The sampling simulations lasted for \SI{100}{\pico\second} and were simulated with a Berendsen thermostat using a coupling constant of \SI{500}{\femto\second} and a time step of \SI{0.5}{\femto\second}.

\begin{figure}
 \centering
 \includegraphics[width=8cm]{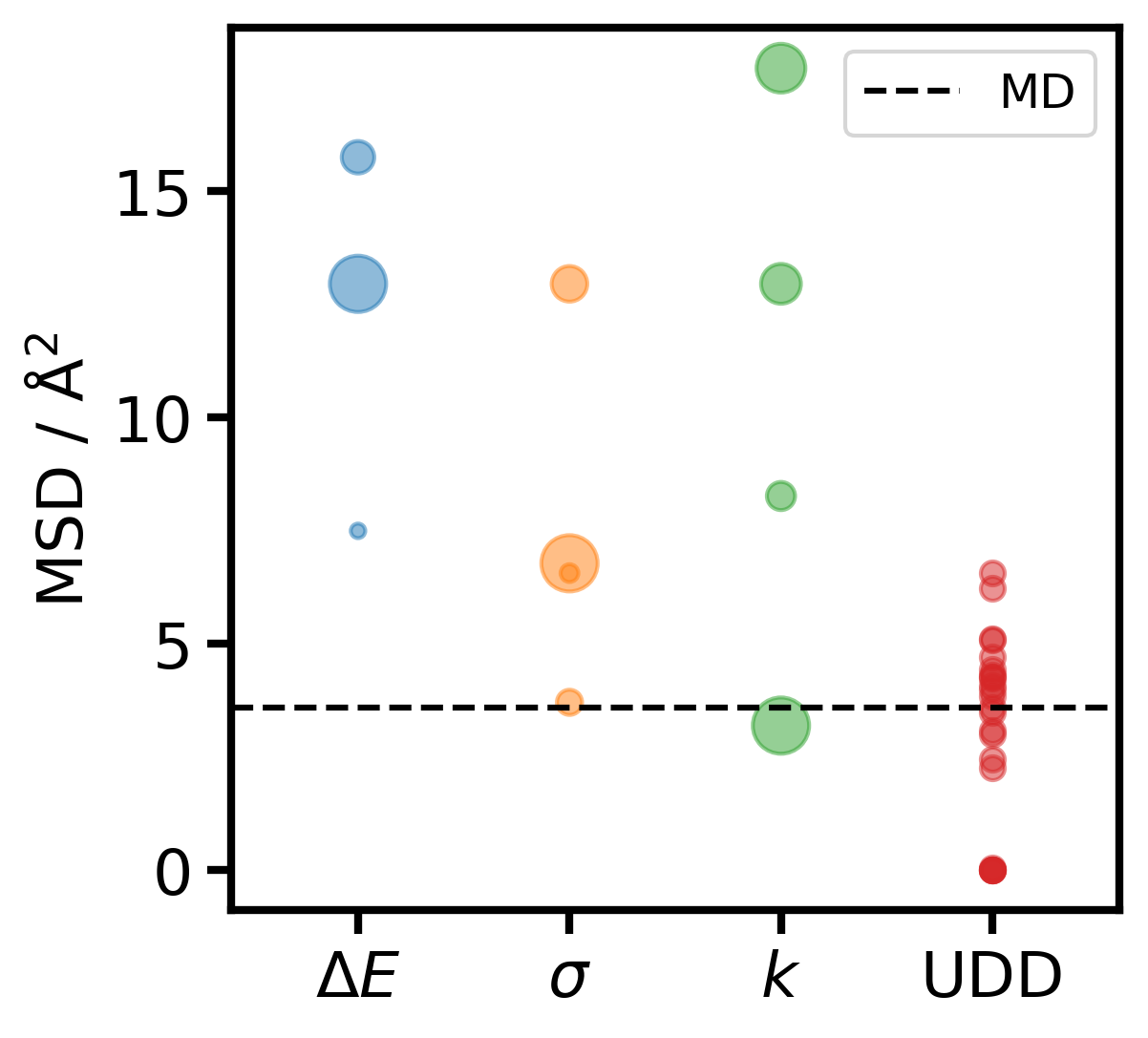} 
 \caption{
    Final mean square displacements \ce{BF4-} center of masses from \ce{BMIM+BF4-}trajectories simulated with MD and various parameters for ERBS and UDD. For the three ERBS parameter scans, the size of the dots is proportional to the value of the respective parameter. For ERBS, these ranges are $\Delta E = \{ 0.1, 0.5, 1.5, 2.5\}$~\si{\electronvolt\per\atom}, $\sigma = \{ 0.5, 1.0, 2.0, 5.0\}$ and $k= \{ 1,2,3,4\}$ and for UDD $A = \{ 0.1, 0.5, 1.0, 2.0, 5.0, 10.0\}$ and $B = \{0.1, 0.5, 1.0, 2.0 \}$. 
 }
 \label{fig:udd}
\end{figure}

To assess sampling efficiency, we use the mean squared displacement (MSD) of the \ce{BF4} center of mass, as defined in \cref{eq:msd}, as a proxy measure.
A higher MSD reflects a larger deviation from the initial configuration, indicating more extensive sampling of the intermolecular degrees of freedom.
The maximum MSD for each trajectory is displayed in \cref{fig:udd}.

It can be seen that UDD is capable of enhancing the sampling of intermolecular degrees of freedom to some degree, but only a few parameter choices offer a significant enhancement.
The trajectory with an MSD of \SI{0.5}{\square\angstrom\per\nano\second} terminated early due to bond breakage, leading to small overall displacements.
All runs with $B=0.1$~eV
terminated within the first 100 simulation steps, before a configuration could be collected.
In the case of ERBS, only the $\Delta E = 2.5$~\si{\electronvolt\per\atom} run is terminated early for the same reason.
We find that for most parameter choices, ERBS drastically enhances the intermolecular motions, while the enhancement for UDD is significantly less pronounced.
In the best cases, ERBS achieves an MSD almost 5 times larger than unbiased MD, while UDD only shows an increase of a factor of 1.8.

\begin{figure}
 \centering
 \includegraphics[width=8cm]{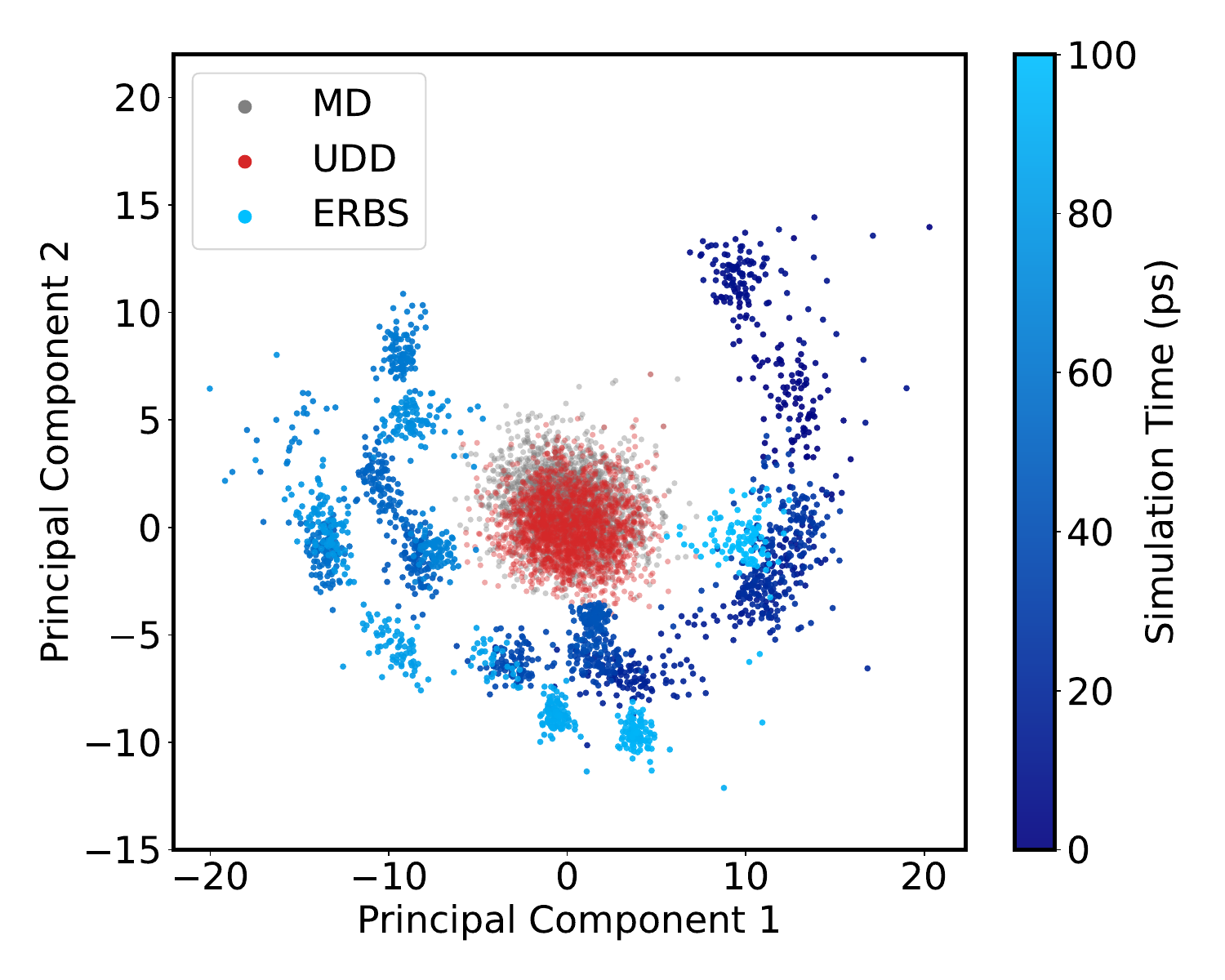} 
 \caption{
    First two principal components of the system average descriptors of \ce{BMIM+BF4-} for an MD, a UDD, and an ERBS trajectory.
    The ERBS descriptors are color-coded by the simulation time to highlight the clustering.
 }
 \label{fig:bmim_pca}
\end{figure}

To visualize the differences in exploration strategies, we consider the distribution of averaged Gaussian moment descriptors.
We compute the average descriptors for the MD trajectory and the highest MSD UDD, and ERBS trajectories, and project them into a common 2D PCA basis.
\Cref{fig:bmim_pca} reveals that the UDD trajectory samples configurations that are shifted with respect to the MD trajectory in the dimensionality-reduced descriptor space.
In contrast, the ERBS trajectory displays a distinct clustering pattern that evolves over time, a direct consequence of the iterative deposition of bias potential in descriptor space.
Overall, the ERBS trajectory exhibits the highest variance in descriptor space, highlighting the method's ability to drive the system out of local minima and explore a significantly larger volume of the physically relevant configurational space.

To explain the sampling behaviour of UDD, we consider the bias forces
\begin{equation}
    -\frac{\partial E_{\text{UDD}}}{\partial \mathbf{R}} = - \frac{\partial E_{\text{UDD}}\left( \sigma_E^2 \right)}{\partial \sigma_E^2}  \cdot \frac{\partial \sigma_E^2}{\partial \mathbf{R}},
\end{equation}
where the derivative of the predicted energy variance $ \sigma_E^2 $ with respect to atomic positions $ \mathbf{R} $ is proportional to
\begin{equation}\label{eq:uddbias}
    \frac{\partial \sigma_E^2}{\partial \mathbf{R}} \propto \sum_m \left( E^{(m)} - \bar{E} \right) \left( \mathbf{F}^{(m)} - \bar{\mathbf{F}} \right).
\end{equation}
As such, the bias forces depend on the disagreement between the ensemble members' force predictions.

In molecular liquids, forces predictions can be analytically decomposed into vibrational, rotational, and translational components\cite{magdauMachineLearningForce2023a,thompsonLAMMPSFlexibleSimulation2022}, with the intermolecular forces typically being an order of magnitude smaller than the intramolecular ones.
Hence, the intermolecular bias forces depend on the disagreement of the ensemble members' intermolecular force predictions.
This decomposition is illustrated in the parity plots shown in \cref{forcedecomp}.
Both the predicted intermolecular forces and their uncertainties remain small for a well-trained and well-calibrated MLIP.
Since the UDD forces along particular degrees of freedom are directly linked to the model's uncertainty along these degrees of freedom, the bias tends to vanish in the directions of slow, collective motion.
This limits the effectiveness of UDD in enhancing sampling along shallow degrees of freedom.

\section{Conclusion}

In this work, we have introduced ERBS, a general-purpose enhanced sampling strategy that allows for the construction of diverse datasets for MLIPs.
Starting from the model descriptor averaged over the whole system, ERBS identifies the slowest collective modes in descriptor space via principal component analysis.
A bias potential is then constructed based on the recently introduced OPES-Explore method, which combines desirable attributes such as rapid exploration of the FES and limiting the maximum strength of the bias potential.
Unlike uncertainty-based enhanced sampling methods, ERBS does not require a preliminary MLIP and can be used on top of any interatomic potential. 
Model independence makes ERBS particularly convenient in early-stage dataset generation or for systems where classical force fields or pre-trained MLIPs are readily available.

We first evaluated ERBS using a classical force field for alanine dipeptide.
ERBS was able to achieve up to \SI{75}{\percent} coverage of the dihedral angle space in just \SI{80}{\pico\second} of simulation, and the resulting data enabled the training of MLIPs that accurately reproduce the free energy landscape.

To demonstrate utility in an active learning setting, we applied ERBS to the iterative construction of a water dataset.
Compared to unbiased sampling, ERBS dramatically accelerated convergence toward accurate diffusion coefficients, matching the quality of models trained on literature datasets using an order of magnitude fewer training points.
The model trained exclusively with MD samples, on the other hand, showed significant deviations from the reference diffusion coefficient throughout.

Finally, we compared the sampling behavior of ERBS and uncertainty-driven dynamics for the highly viscous ionic liquid \ce{BMIM+BF4-}.
We observed that ERBS more effectively enhances sampling along slow, intermolecular degrees of freedom.
The accelerated sampling is attributed to the use of global, low-dimensional collective variables, which avoids overemphasising high-frequency intramolecular modes, which are often the largest contributors to mode uncertainty and error.
ERBS offers a fast method for enhancing the quality of training data both in static and active learning settings, even when no pre-trained model is available.
Crucially, although demonstrated here using the Gaussian Moment descriptor, the framework is representation-agnostic and can be directly coupled to other state-of-the-art descriptors or the feature layers of equivariant neural networks.
The role of a pre-trained representation and extensions to constant-pressure simulations will be explored in subsequent studies.
While this study focuses on the exploration of molecular and liquid configuration spaces, the variance-based identification of collective variables is a phase-agnostic principle, suggesting that the method is applicable to solid-state sampling as well.
We intend to further investigate this line of research in future work.

Lastly, the use of enhanced sampling techniques, such as ERBS, may also prove valuable in the context of constructing datasets for atomistic foundation models.
By systematically exploring underrepresented regions of configuration space, these methods can help ensure broad coverage of structural motifs.
This could lead to more compact and diverse datasets, ultimately improving the transferability and robustness of foundation models across domains.

\section*{Associated Content}
\subsection*{Data Availability}
The workflow notebook as well as all input files for the various software packages needed to reproduce the work presented here can be found at \url{https://github.com/M-R-Schaefer/erbs_experiments/}.
All data generated during the iterative training and production simulations are stored on an S3-object storage.
It can be obtained by cloning the repository and executing \texttt{dvc pull} in the repository folder.

\subsection*{Code Availability}
All software used throughout this work is publicly available.
ERBS is available on Github at \url{https://github.com/apax-hub/erbs}.
The Apax repository is available on Github at \url{https://github.com/apax-hub/apax}.
IPSuite is available at \url{https://github.com/zincware/IPSuite}.
All three can be installed from PyPi \textit{via} \texttt{pip install erbs apax ipsuite}.

\subsection*{Supporting Information}

Detailed table of model hyperparameters for all experiments; discussion of the computational complexity of the method; Ramachandran space coverage for various models of the alanine dipeptide experiment; calibration metrics of the shallow ensemble in the BMIM experiment; decomposition of forces into translational, rotational and vibrational components for the BMIM dataset; test set prediction-error parity plots for all experiments.

\section*{Author Information}

\subsection*{Corresponding Author}
\textbf{Johannes K\"{a}stner} - Institute for Theoretical Chemistry, University of Stuttgart, Pfaffenwaldring 55, 70569 Stuttgart, Germany; https://orcid.org/0000-0001-6178-7669; Email: kaestner@theochem.uni-stuttgart.de

\subsection*{Author}
\textbf{Moritz R. Sch\"{a}fer} - Institute for Theoretical Chemistry, University of Stuttgart, Pfaffenwaldring 55, 70569 Stuttgart, Germany; https://orcid.org/0000-0001-8474-5808

\subsection*{Notes}
The authors have no conflicts of interest to declare.

\section*{Acknowledgements}
The authors would like to thank Nico Segreto and Fabian Zills for insightful discussion and
comments on an early version of the manuscript.
J.K., and M.S. acknowledge support by the Deutsche Forschungsgemeinschaft (DFG, German Research Foundation) in the framework of the priority program SPP 2363, “Utilization and Development of Machine Learning for Molecular Applications - Molecular Machine Learning” Project No. 497249646.
Further funding was provided by Deutsche Forschungsgemeinschaft (DFG, German Research Foundation) under Germany's Excellence Strategy - EXC 2075 – 390740016. We acknowledge the support by the Stuttgart Center for Simulation Science (SimTech).

All authors acknowledge support by the state of Baden-Württemberg through bwHPC
and the German Research Foundation (DFG) through grant INST 35/1597-1 FUGG.

%% file: supplementary-information.tex
\begin{suppinfo}

\setcounter{table}{0}
\setcounter{figure}{0}
\setcounter{equation}{0}
\setcounter{section}{0}

\renewcommand*{\thepage}{S\arabic{page}}
\renewcommand{\theequation}{S\arabic{equation}}
\renewcommand{\thetable}{S\arabic{table}}
\renewcommand{\thefigure}{S\arabic{figure}}
\renewcommand{\thesection}{S\arabic{section}}
\renewcommand{\thelstlisting}{S\arabic{lstlisting}}

\section{Model Hyperparameters}\label{si:model_hypers}
All hyperparameters used to train the models of each experiment are listed in \cref{tab:si:hypers}.
The Huber loss function used for the alanine dipeptide experiment is given by

\begin{equation}\label{si:eq:huber}
    \mathcal{L}_\delta(\boldsymbol{\theta}) =
    \sum_{k=1}^{N_{\text{train}}} \sum_i^{ N_{ \text{atoms} }^{(k)}} \frac{1}{3 N_{\text{atoms}}^{(k)} }
    \begin{cases}
    \frac{1}{2}(\boldsymbol{F}^{\text{ref}}_{i,k} - \boldsymbol{F}_i(S_k, \boldsymbol{\theta}))^2, & \text{if } |\boldsymbol{F}^{\text{ref}}_{i,k} - \boldsymbol{F}_i(S_k, \boldsymbol{\theta})| \leq \delta \\
    \delta \left( |\boldsymbol{F}^{\text{ref}}_{i,k} - \boldsymbol{F}_i(S_k, \boldsymbol{\theta})| - \frac{1}{2} \delta \right), & \text{otherwise}
    \end{cases}
\end{equation}

In total, the number of trainable parameters was 32929 for the Alanine Dipeptide, 90244 for water, and 92548 \ce{BMIM+BF4-} experiments.
The difference in parameter counts for the last two experiments arises from the different number of elements in the systems, which results in a different number of elemental embedding parameters.

\begin{table}[ht]
\caption{GMNN training hyperparameters used throughout the experiments of Alanine Dipeptide, Water and \ce{BMIM+BF4-}}
\label{tab:si:hypers}
\centering
\small
    \begin{tabular}{@{}lll@{}}
    \toprule
        \textbf{Hyperparameter} & \textbf{Alanine Dipeptide} & \textbf{Water, \ce{BMIM+BF4-}} \\
        \midrule
        \multicolumn{3}{l}{\textit{Training}} \\
        \midrule
            Epochs & 10\,000 & 10\,000 \\
            Batch size & 8 & 1 \\
            Gradient clipping & 10.0 & 10.0 \\
        \midrule
        \multicolumn{3}{l}{\textit{Model}} \\
        \midrule
            Basis function & Bessel, $n=16$, $r_\text{max}=5.0$ & Bessel, $n=16$, $r_\text{max}=5.5$ \\
            Radial functions & 5 & 6 \\
            NN layers & 64, 64 & 128, 64 \\
            Ensemble & --- & Shallow (16 members) \\
        \midrule
        \multicolumn{3}{l}{\textit{Optimizer}} \\
        \midrule
            Name & AdamW & AdamW \\
            Embedding LR & 0.001 & 0.0001 \\
            NN LR & 0.001 & 0.0001 \\
            Scale LR & 0.0005 & 0.0001 \\
            Shift LR & 0.0005 & 0.0001 \\
            Weight decay & $10^{-5}$ & $2 \cdot 10^{-4}$ \\
        \midrule
        \multicolumn{3}{l}{\textit{Schedule}} \\
        \midrule
            Name & Cyclic cosine & Cyclic cosine \\
            Period & 50 & 50 \\
            Decay factor & 0.95 & 0.96 \\
        \midrule
        \multicolumn{3}{l}{\textit{Loss functions}} \\
        \midrule
            Energy loss & Huber ($\delta=0.5$), weight=1.0 & NLL, weight=1.0 \\
            Force loss & Huber ($\delta=0.1$), weight=2.0 & NLL, weight=2.0 \\
    \bottomrule
    \end{tabular}

\end{table}

\section{Scaling}
The method by \citeauthor{yooMetadynamicsSamplingAtomic2021} constructs per-atom bias potentials based on Behler-Parinello descriptors and a metadynamics-like functional form of the bias.
The use of uncompressed per-atom descriptors significantly increases the memory requirements for storing the reference descriptors.
Further, the calculation of $E_{\text{bias}}$ requires comparison of each atom's descriptor with all reference descriptors of the same element.
This introduces an additional factor to the cost calculation of the bias potential of $N_{\text{atoms}}$.
ERBS uses diagonal kernels; the approach of \citeauthor{yooMetadynamicsSamplingAtomic2021}\cite{yooMetadynamicsSamplingAtomic2021} constructs an adaptive covariance matrix from descriptor Jacobians according to:

 \begin{equation}
    \Sigma_{jk}(\mathbf{G}) = \sigma^2 \sum_{i=1}^{N_\text{at}} \sum_{\alpha = x,y,z} \frac{\partial G_j}{\partial R_{i,\alpha}} \frac{\partial G_k}{\partial R_{i,\alpha}} + \varepsilon \delta_{jk},
\end{equation}

where $\sigma$ is a hyperparameter controlling the scale of the kernel, and $\epsilon$ is a regularization constant added to the diagonal to ensure numerical stability.
The covariance matrix has a dimensionality of $N_{\text{feat}} \times N_{\text{feat}}$, where $N_{\text{feat}}$ is the number of descriptor components per configuration.
Its inverse needs to be stored along with the reference descriptors or recomputed on the fly, adding to either the memory or computational cost.

Additionally, we find that, at least for the GM descriptor, this matrix can be severely ill-conditioned, such that its inverse is dominated by the choice of $\epsilon$, effectively suppressing the geometry-dependent structure of the covariance.
One possible cause for the ill-conditioning is that the GM descriptor constructs many body features via outer products, leading to high correlation of the features.
The Behler-Parinello descriptor, on the other hand, has separate parts for two and three-body interactions of element pairs and triplets, reducing the correlation.

\section{Ramachandran Space Coverage}\label{coverage}

To further analyze the exploration behavior of the ERBS method in comparison to high-temperature molecular dynamics, we investigate the Ramachandran space coverage visually.
\Cref{fig:si:coverage_si} displays the coverage plots for the MD 1200 K, ERBS A, ERBS B trajectories, and one that led to the dissociation of the molecule.
The parameter choice for the dissociated simulation was $\Delta E = 10$ eV,  $\sigma=0.1$, and $k=2$.
We find that the MD simulation at 1200 K does not sample the free energy minimum in the lower right quadrant, while both ERBS simulations do.
The good description of the minimum location and high errors on the right-hand side of the FES of the MD 1200 K model can be explained in this way.
The broken simulation achieves a high coverage, although this is merely due to the free rotation of fragments.

\begin{figure}
 \centering
 \includegraphics[width=0.8\linewidth]{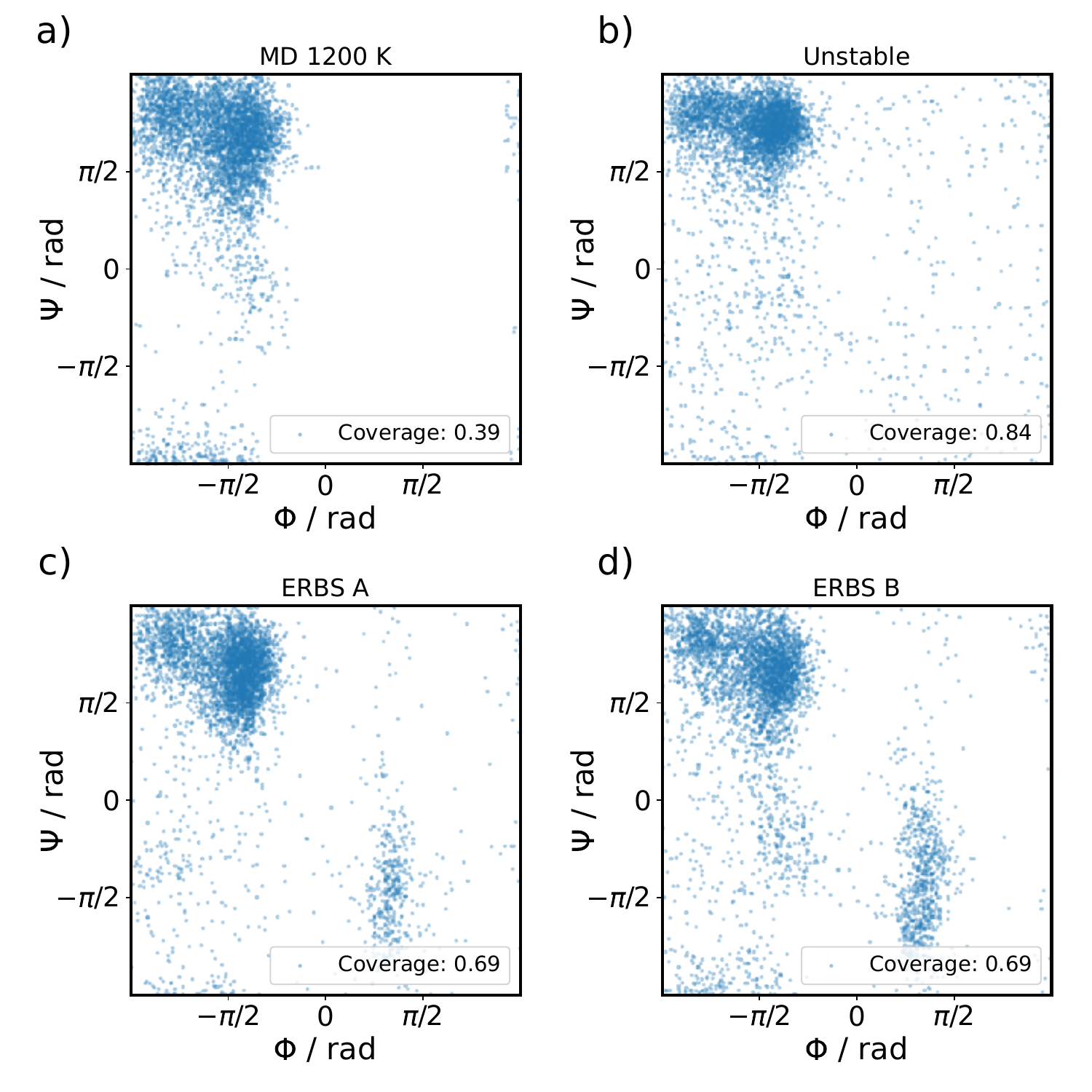} 
 \caption{
    Ramachandran space coverage of the a) MD 1200K, b) $\Delta E = 10$ eV,  $\sigma=0.1$, and $k=2$, c) ERBS A, and d) ERBS B.
 }
 \label{fig:si:coverage_si}
\end{figure}

\section{Radial Distribution Functions of Water}\label{rdf}

We further investigate the capability of the active learned MLIP to reproduce structural properties of liquid water.
Specifically, we calculate the oxygen-oxygen radial distribution functions (RDFs).
As shown in \Cref{fig:si:water_rdf}, the RDFs for all models are in good agreement and visually indistinguishable, accurately reproducing the location and intensity of both the first and second hydration shells.

\begin{figure}
 \centering
 \includegraphics[width=0.5\linewidth]{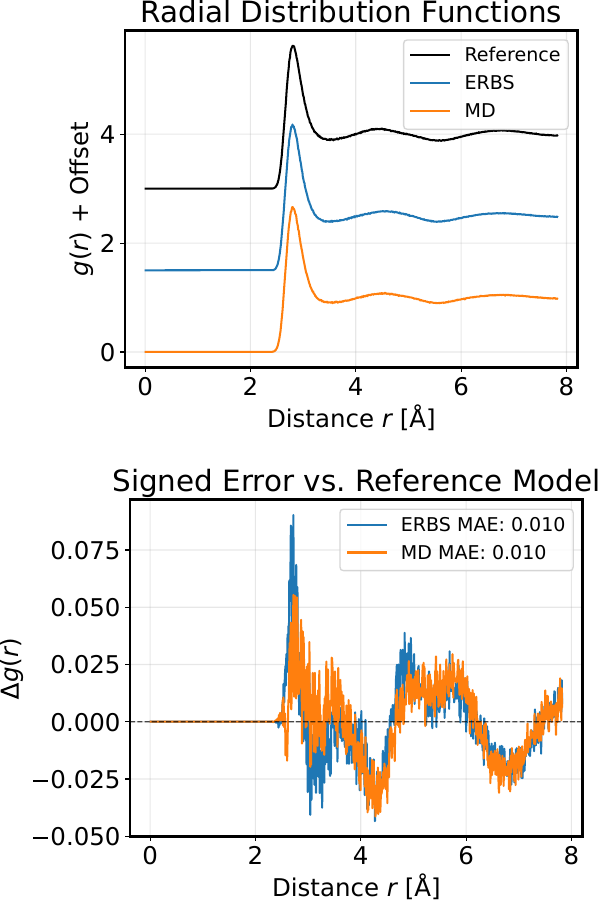} 
 \caption{
    Oxygen-oxygen radial distribution functions of the reference model and the MLIPs produced by active learning on MD and ERBS data. The right panel shows the signed error of the active learned models with respect to the reference.
 }
 \label{fig:si:water_rdf}
\end{figure}

We do not observe an underestimation of the hydration shells that would explain an overestimated diffusion coefficient.
The radial distribution function defines the effective 2-body potential of mean force (PMF), $W(r) = - k_{\text{B}} T \log g(r)$.
However, $W(r)$ is a free energy surface resulting from integrating out all other degrees of freedom, including higher body order terms.

Diffusion, however, is a dynamical process governed by the underlying high-dimensional potential energy surface, not the projected PMF.
A transition event often requires a specific many-body rearrangement, such as a rotation around a bond that constitutes a high-energy barrier on the true PES.

The fact that the MD-trained model reproduces the RDF, and thus the potential of mean force, but overestimates diffusion implies it has learned a 'flattened' PES.
It captures the pairwise energetics correctly but lacks the explicit many-body repulsions that create the true friction and transition barriers in the liquid.

\section{Calibration Metrics}\label{calmetrics}

Assessing the quality of uncertainty estimates is crucial in the active learning context, both for terminating sampling trajectories and for the appropriate biasing by UDD.
\Cref{fig:si:calibration} displays the scatter plots for predicted uncertainties compared to empirical errors for energy and force uncertainties for shallow ensembles trained with MSE and NLL losses.
While the validation errors are comparable for both cases, the calibration of uncertainty estimates is significantly improved by the use of the probabilistic loss function.

\begin{figure}
 \centering
 \includegraphics[width=0.9\linewidth]{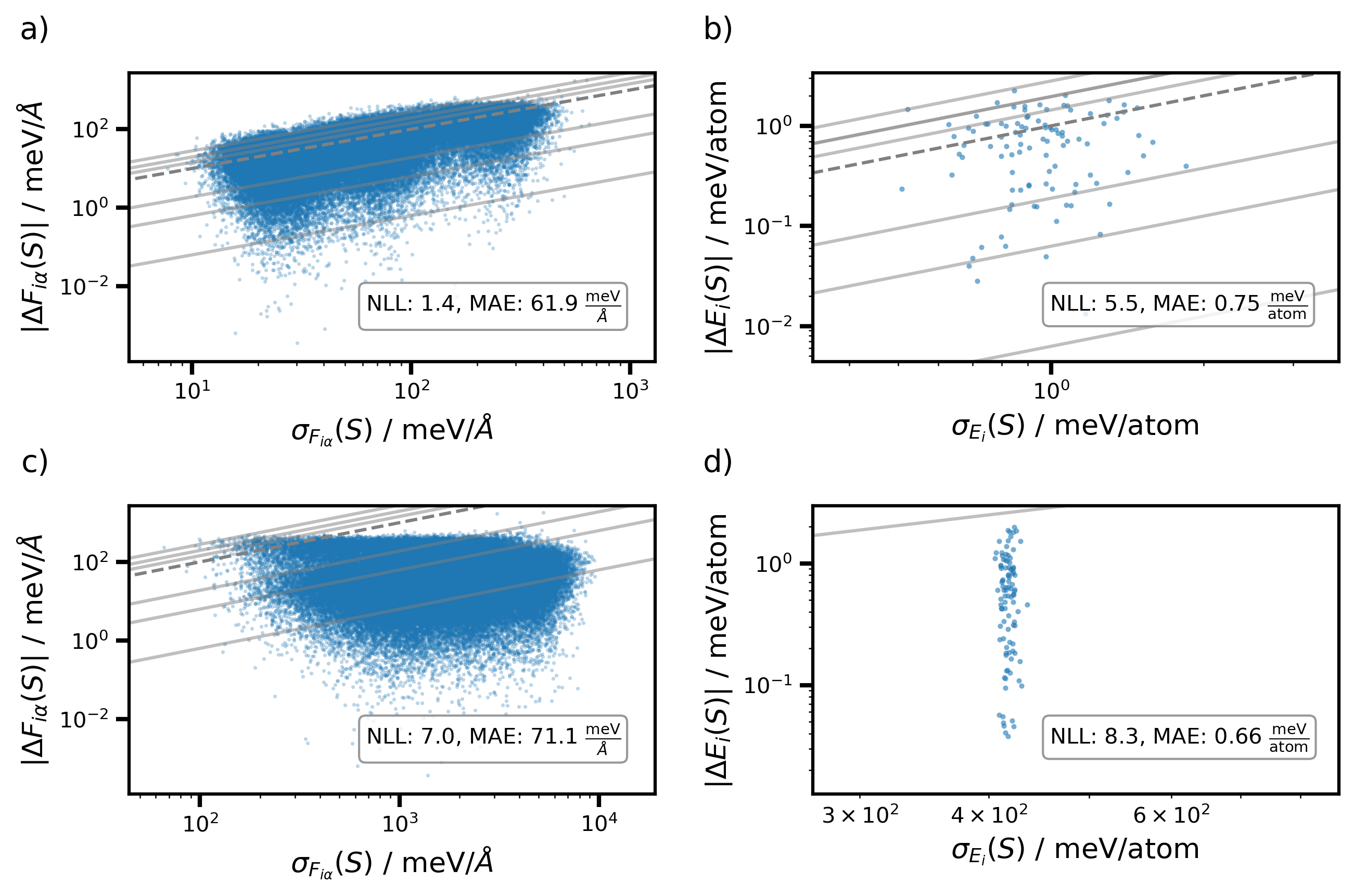} 
 \caption{
    Predicted-empirical error scatter plot for energies and forces of the \ce{BMIM+BF4-} validation dataset. Panels a) and b) show the performance of the shallow ensemble trained with an NLL loss, c) and d) that of the shallow ensemble trained with an MSE loss.
 }
 \label{fig:si:calibration}
\end{figure}

\section{Force Decomposition}\label{forcedecomp}

In molecular liquids, the forces on each atom can be analytically decomposed into vibrational, translational, and rotational forces\cite{magdauMachineLearningForce2023a}.
\Cref{fig:si:forcedecomp} shows the prediction errors of the model trained on \ce{BMIM+BF4-} for the decomposed forces.
The separation in magnitude between inter- and intramolecular forces is evident.
While the MAE for the intermolecular forces is fairly small, so are the true forces.
The mean absolute force components of the translational and rotational are \SI{20.6}{\milli\electronvolt\per\angstrom} and \SI{19.5}{\milli\electronvolt\per\angstrom} respectively,  resulting in large relative errors.

\begin{figure}
 \centering
 \includegraphics[width=0.9\linewidth]{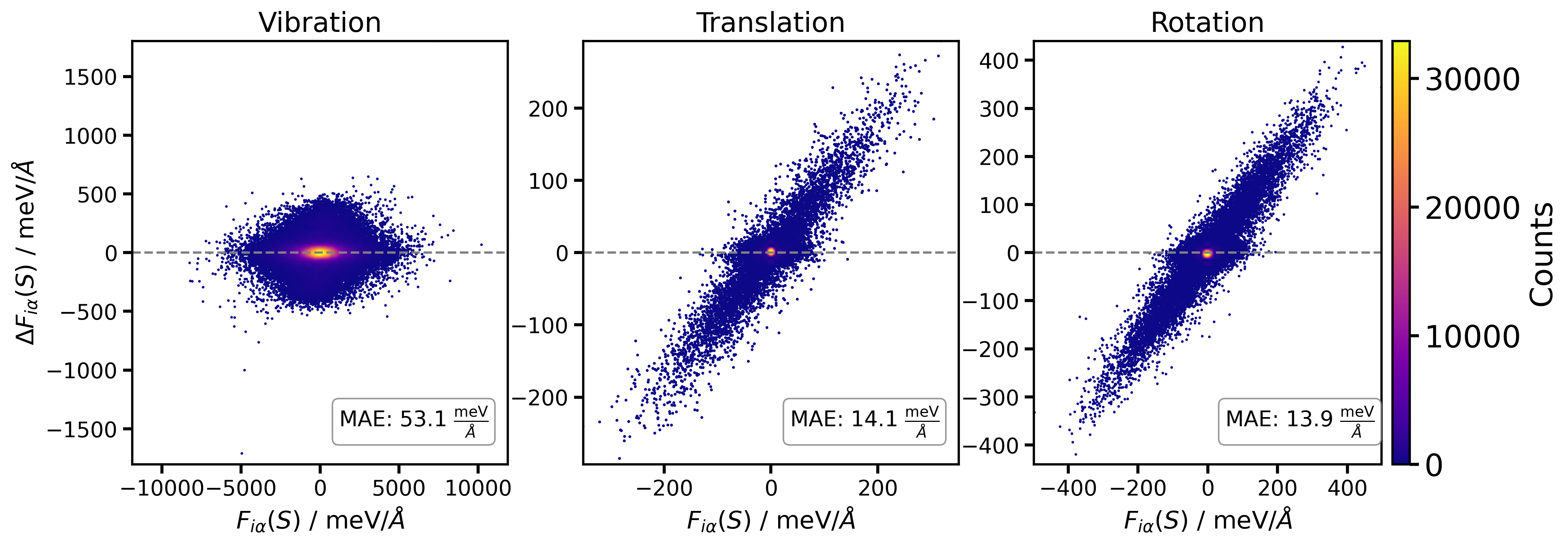} 
 \caption{
    Error of predicted compared to true force components for the \ce{BMIM+BF4-} validation dataset.
 }
 \label{fig:si:forcedecomp}
\end{figure}

\section{Test Set Error Analysis}\label{parity}

To evaluate predictive quality, we present density-colored parity plots of energy and force errors for key models. For the alanine dipeptide, where cross-validation results are detailed in the main text, \cref{fig:si:parity_alad} shows the self-validation performance. As the reference data is derived from a classical force field, all models fit energies and forces well below chemical accuracy.

For the water experiment, we compare the final MD and ERBS active learning models against the model trained on the literature dataset. While the MD and ERBS models achieve similar high accuracy, the literature-trained model exhibits consistently higher errors. This is attributed to the broader diversity of the literature dataset, which includes path-integral MD and constant-pressure simulations.

Finally, for \ce{BMIM+BF4-}, we analyze the shallow ensemble trained on the literature dataset. It achieves improved energy metrics compared to the original publication (MAE: \SI{5.5}{\milli\electronvolt\per\atom}), though force metrics are slightly higher (reference MAE: \SI{48}{\milli\electronvolt\per\angstrom}).

\begin{figure}
 \centering
 \includegraphics[width=8cm]{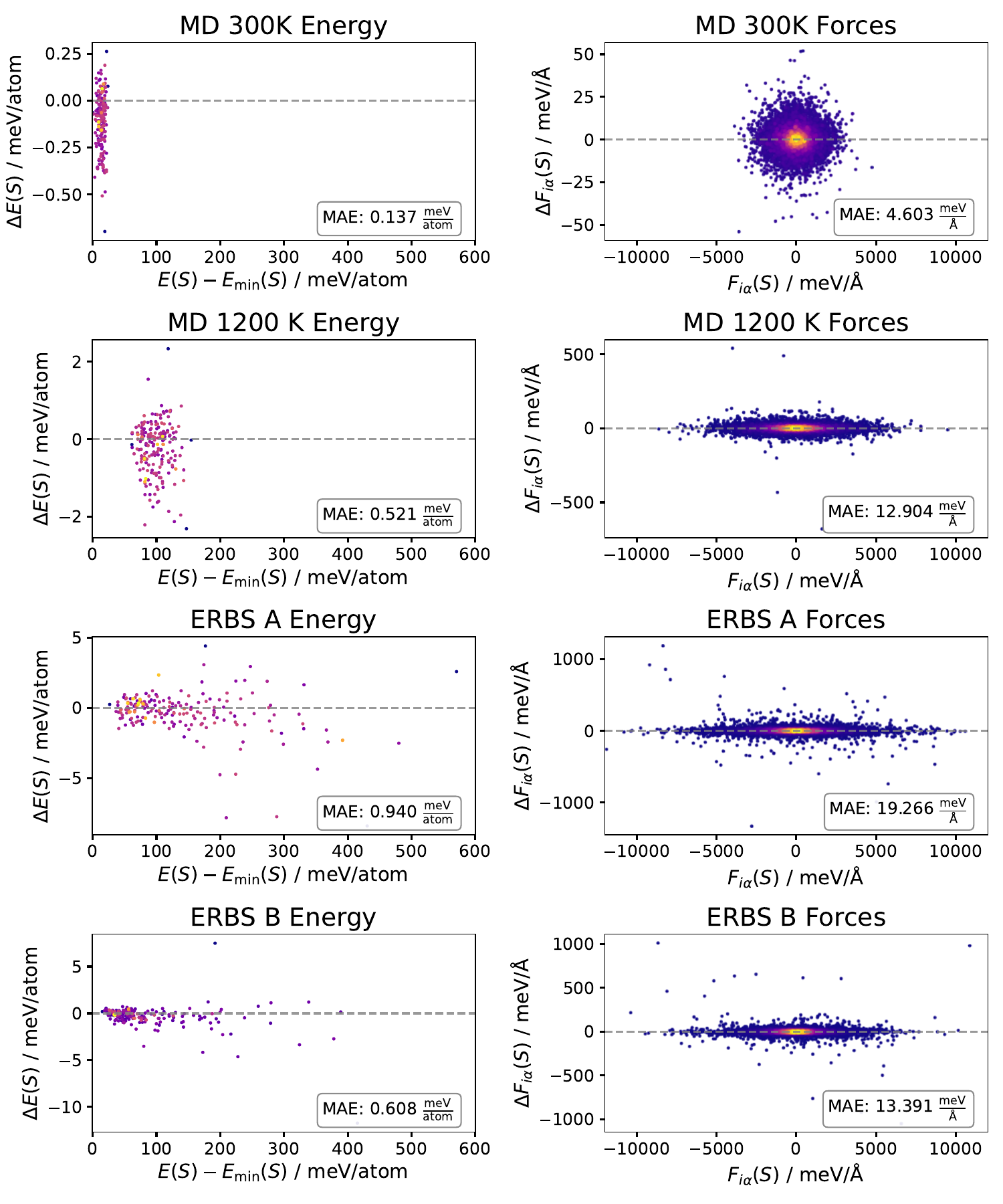} 
 \caption{
    Energy and force prediction errors of the MD 300 K, MD 1200 K, ERBS A and ERBS B on their respective validation sets.
 }
 \label{fig:si:parity_alad}
\end{figure}

\begin{figure}
 \centering
 \includegraphics[width=8cm]{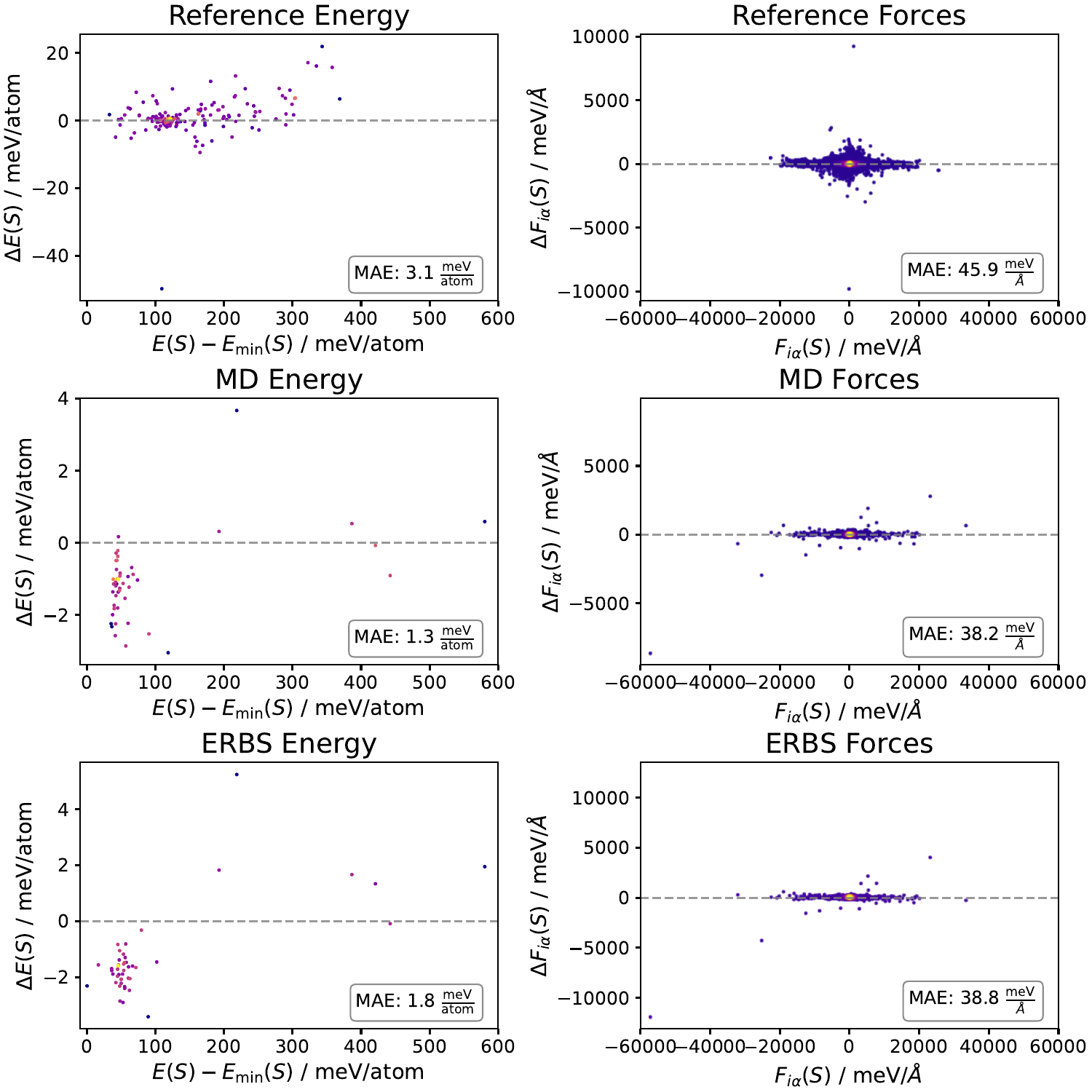} 
 \caption{
    Energy and force prediction errors of the reference model and the active learned models trained on MD and ERBS data.
 }
 \label{fig:si:parity_water}
\end{figure}

\begin{figure}
 \centering
 \includegraphics[width=8cm]{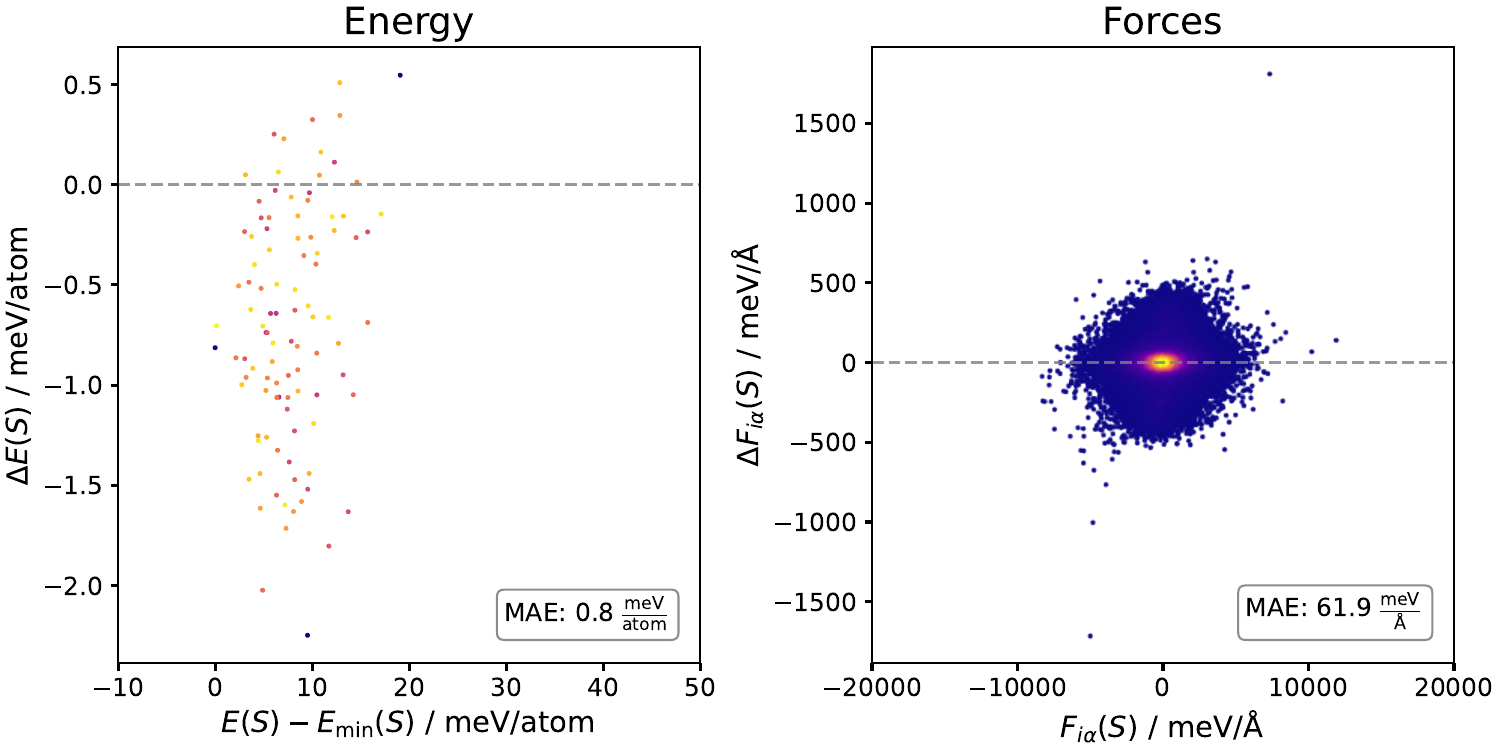} 
 \caption{
    Energy and force prediction errors of the model trained on the \ce{BMIM+BF4-} literature dataset.
 }
 \label{fig:si:parity_bmim}
\end{figure}

\end{suppinfo}